\documentclass{amsproc}

\usepackage{latexsym}
\usepackage{amsmath}
\usepackage{bm}
\usepackage{amssymb}
\usepackage{amsfonts}
\usepackage{amsthm}
\usepackage{cite}
\usepackage{fullpage}
\usepackage{graphicx}
\usepackage{subfig}
\usepackage{tikz,tikz-qtree,tikz-qtree-compat}
\usetikzlibrary{shapes,snakes,arrows,calc,fit}
\usepackage{color,xspace,colortbl}
\usepackage{float}
\usepackage[ruled]{algorithm2e}
\usepackage[noend]{algpseudocode}
\usepackage{enumitem}
\usepackage{balance}
\usepackage{todonotes}
\usepackage{url}
\usepackage{multirow}

\colorlet{LightViolet}{violet!40}
\colorlet{LightRed}{red!40}
\colorlet{LightOrange}{orange!40}
\colorlet{LightGreen}{green!40}
\colorlet{LightBlue}{blue!40}
\colorlet{DarkGreen}{green!50!black}
\colorlet{DarkRed}{red!70!black}
\colorlet{DarkCyan}{red!70!black}
\colorlet{DarkBlue}{blue!80!black}
{\definecolor{DarkOrange}{rgb}{1.0, 0.49, 0.0}
\definecolor{Airforceblue}{rgb}{0.36, 0.54, 0.66}

\newcommand{\nop}[1]{}

\newcommand{\heavy}{\text{\sf heavy}}
\newcommand{\light}{\text{\sf light}}
\newcommand{\fecp}{\text{\sf FECP}}
\newcommand{\poly}{\text{\sf poly}}
\newcommand{\dc}{\text{\sf DC}}

\newcommand{\FD}{\text{\sf FD}}
\newcommand{\hdc}{\text{\sf HDC}}

\newcommand{\incomp}{\perp}
\newcommand{\flow}{{\sf inflow}}

\newcommand{\LB}{\text{{\sf LogicBlox}}}
\newcommand{\lftj}{\text{{\sf LFTJ}}}
\newcommand{\gj}{\text{{\sf Generic-Join}}}

\newcommand{\csma}{\text{\sf CSMA}}
\newcommand{\panda}{\text{\sf PANDA}}

\newcommand{\complexityclass}{\mathbf}

\newcommand{\np}{\complexityclass{NP}}

\newcommand{\functionname}[1]{\text{\sf #1}}

\newcommand{\dom}{\functionname{Dom}}

\newcommand{\wcoj}{\functionname{WCOJ}}

\newcommand{\nprr}{\text{\sf NPRR}\xspace}

\newcommand{\agm}{\text{\sf AGM}\xspace}

\newcommand{\pr}{\mathbb P}
\newcommand{\supp}{\text{\sf supp}}

\newcommand{\calP}{\mathcal P}

\newcommand{\calD}{\mathcal D}
\newcommand{\calE}{\mathcal E}

\newcommand{\calG}{\mathcal G}
\newcommand{\calH}{\mathcal H}
\newcommand{\calV}{\mathcal V}

\newcommand{\N}{\mathbb N} % the natural numbers
\newcommand{\R}{\mathbb R} % the real numbers
\newcommand{\Q}{\mathbb Q} % the real numbers
\newcommand{\Mod}{\text{\sf M}}                                                 
\newcommand{\sa}{\text{\sf SA}}

\newcommand{\be}{\begin{enumerate}}
\newcommand{\ee}{\end{enumerate}}
\newcommand{\bi}{\begin{itemize}}
\newcommand{\ei}{\end{itemize}}
\newcommand{\beq}{\begin{equation}}
\newcommand{\eeq}{\end{equation}}

\newcommand{\bp}{\begin{proof}}
\newcommand{\ep}{\end{proof}}
\newcommand{\bcor}{\begin{cor}}
\newcommand{\ecor}{\end{cor}}
\newcommand{\bthm}{\begin{thm}}
\newcommand{\ethm}{\end{thm}}
\newcommand{\blmm}{\begin{lmm}}
\newcommand{\elmm}{\end{lmm}}
\newcommand{\bdefn}{\begin{defn}}
\newcommand{\edefn}{\end{defn}}
\newcommand{\bprop}{\begin{prop}}
\newcommand{\eprop}{\end{prop}}
\newcommand{\bconj}{\begin{conj}}
\newcommand{\econj}{\end{conj}}
\newcommand{\bopm}{\begin{opm}}
\newcommand{\eopm}{\end{opm}}
\newcommand{\brmk}{\begin{rmk}}
\newcommand{\ermk}{\end{rmk}}

\newcommand{\suchthat}{\ | \ }
\newcommand{\inner}[1]{\left\langle #1 \right\rangle}

\newcommand{\mv}[1]{\mathbf{#1}}

\theoremstyle{plain}                   % default
\newtheorem{thm}{Theorem}[section]
\newtheorem{lmm}[thm]{Lemma}
\newtheorem{prop}[thm]{Proposition}
\newtheorem{cor}[thm]{Corollary}

\theoremstyle{definition}              % Examples and all

\newtheorem{opm}{Open Problem}
\newtheorem{conj}{Conjecture}
\newtheorem{ex}{Example}

\newtheorem{defn}{Definition}

\newtheorem{rmk}{Remark}

\definecolor{Red}{RGB}{255,204,204}
\definecolor{Green}{RGB}{204,255,204}
\definecolor{Blue}{RGB}{204,204,255}

\usepackage{xpatch}
\usepackage{textcase}
\makeatletter
\xpatchcmd{\@sect}{\uppercase}{\MakeTextUppercase}{}{}
\xpatchcmd{\@sect}{\uppercase}{\MakeTextUppercase}{}{}
\makeatother

\allowdisplaybreaks[1]

\usepackage[colorlinks=true,linkcolor=blue,citecolor=blue]{hyperref}

\usepackage{booktabs} % For formal tables

\title{Worst-Case Optimal Join Algorithms: \\Techniques, Results, and Open Problems}

\author{
   {\sf Hung Q. Ngo}\\
   RelationalAI, Inc.
}

\begin{document}

\maketitle
\begin{abstract}
   Worst-case optimal join algorithms are the class of join algorithms whose 
   runtime match the worst-case output size of a given join query. While the 
   first provably worst-case optimal join algorithm was discovered relatively 
   recently, the techniques and results surrounding these algorithms grow out 
   of decades of research from a wide range of areas, intimately connecting 
   graph theory, algorithms, information theory, constraint satisfaction, 
   database theory, and geometric inequalities. These ideas are not just 
   paperware: in addition to academic project implementations, two variations of 
   such algorithms are the work-horse join algorithms of 
   commercial database and data analytics engines.

   This paper aims to be a brief introduction to the design and analysis of
   worst-case optimal join algorithms.
   We discuss the key techniques for proving runtime and output size bounds. 
   We particularly focus on the fascinating connection between join algorithms
   and information theoretic inequalities, and the idea of how one can turn a
   proof into an algorithm.
   Finally, we conclude with a representative list of fundamental open problems
   in this area.
\end{abstract}

\section{Introduction}
\label{sec:introduction}

\subsection{Overview}

Relational database query evaluation is one of the most well-studied problems
in computer science. Theoretically, even special cases of the problem are
already equivalent to fundamental problems in other areas; 
for example, queries on {\em one} edge relation can already express various graph
problems, conjunctive query evaluation is deeply rooted in finite model theory
and constraint
satisfaction~\cite{
DBLP:conf/stoc/ChandraM77,
DBLP:journals/jcss/KolaitisV00,
DBLP:journals/jacm/Fagin83,                    
DBLP:conf/stoc/Vardi82,
DBLP:conf/pods/PapadimitriouY97},
and the aggregation version is inference in discrete graphical
models~\cite{faq}.
Practically, relational database management systems (RDBMS) are ubiquitous and
commercially very successful,
with almost 50 years of finely-tuned query evaluation algorithms and 
heuristics~\cite{Selinger:1979:APS:582095.582099,graefe93,DBLP:books/others/red}.

In the last decade or so there have emerged fundamentally new ideas on the three
key problems of a relational database engine: (1) constructing query plans,
(2) bounding intermediate or output size, and (3) evaluating (intermediate)
queries.
The new query plans are based on variable elimination and equivalently tree 
decompositions~\cite{faq,juggling,DBLP:conf/pods/GottlobGLS16}.
The new (tight) size bounds are
information-theoretic, taking into account in a
principled way input statistics and functional
dependencies~\cite{GM06,AGM,GLVV,csma,panda}.
The new algorithms evaluate the multiway join operator in a {\em worst-case
optimal} manner~\cite{skew,LFTJ,nprr,csma,panda},
which is provably asymptotically better than the one-pair-at-a-time join
paradigm.

These fresh developments are exciting both from the theory and from the practical
stand point.  On the theory side, these results
demonstrate beautiful synergy and interplay of ideas from many different research areas:
algorithms, extremal combinatorics, parameterized complexity, information theory, databases,
machine learning, and constraint satisfaction. 
We will briefly mention some of these connections in Sec.~\ref{subsec:history} below.
On the practice side, these results offer their assistance 
``just in time'' for the ever demanding modern data analytics workloads. The 
generality and asymptotic complexity advantage of these algorithms open wider 
the pandora box of true ``in-database'' graph processing, machine learning, 
large-scale inference, and constraint
solving~\cite{DBLP:conf/amw/0001NOS17,DBLP:journals/corr/NgoNOS17,acdc,DBLP:conf/cidr/ElgamalLBETRS17,DBLP:conf/sigmod/FengKRR12,DBLP:journals/pvldb/HellersteinRSWFGNWFLK12,DBLP:conf/sigmod/ArefCGKOPVW15,DBLP:journals/dpd/OlteanuR17,DBLP:journals/pvldb/OlteanuS16,DBLP:conf/sigmod/SchleichOC16,DBLP:conf/sigmod/KumarNP15,DBLP:conf/sigmod/HalperinACCKMORWWXBHS14}.

The reader is referred to~\cite{faq,juggling} for descriptions of the generality
of the types of queries the new style of query plans can help answer. 
In particular, one should keep in mind that the bounds and algorithms described
in this paper apply to aggregate queries in a very general setting, of which
conjunctive queries form a special case. 
The focus of this paper is on the other two developments: output size bounds and
{\em worst-case optimal join} ($\wcoj$) algorithms.

Roughly speaking, a $\wcoj$ algorithm is a join algorithm evaluating a full
conjunctive query in time that is proportional to the worst-case output size of
the query. More precisely, we are given a query $Q$ along with a set $\dc$ of
``constraints'' the input database $\bm D$ is promised to satisfy. The simplest
form of constraints contain the sizes of input relations; these are called {\em
cardinality constraints}. The second form of constraints is prevalent in
RDBMSs, that of functional dependencies ($\FD$). We shall refer to them as 
{\em $\FD$ constraints}. These constraints say that, if we fix the bindings 
of a set $X$ of variables, then
there is at most {\em one} binding for every variable in another set $Y$. 
More generally, there are {\em degree constraints}, which guarantee that
for any fixed binding of variables in $X$, there are at most some given number
of bindings of variables in $Y$. Degree constraints generalize both cardinality
and $\FD$ constraints, because cardinality constraints correspond to degree
constraints when $X = \emptyset$.

We write $\bm D \models \dc$ to denote the fact that the database $\bm D$
satisfies the degree constraints $\dc$.
The worst-case output size problem is to determine the quantity
\begin{align}
   (\text{worst-case output size}) &&
   \sup_{\bm D \models \dc} |Q(\bm D)| \label{eqn:wcos} 
\end{align} 
and a
$\wcoj$ algorithm runs in time $\tilde O(|\bm D| + \sup_{\bm D\models
\dc}|Q(\bm D)|)$, where $\tilde O$ hides a $\log$ factor in the data size and
some query-size dependent factor.
In what follows we present a brief overview of the history of results on
determining~\eqref{eqn:wcos} and on associated $\wcoj$
algorithms. 

Independent of $\wcoj$ algorithms,
the role of bounding and estimating the output size in query
optimizer is of great importance, as estimation errors propagate and sometimes
are as good as random guesses, leading to bad query plans~\cite{DBLP:conf/sigmod/IoannidisC91}.
Hence, as we enrich the class of constraints allowable in the $\dc$ set (say
from upper degree bounds to histogram information or more generally various
statistical properties of the input), one should expect the problem of
determining~\eqref{eqn:wcos} or its expectation to gain more prominence in any
RDBMS.

The role of determining and computing~\eqref{eqn:wcos} in the design of 
$\wcoj$ algorithms, on the other hand, has a fascinating and different bent:
we can turn a mathematical proof of a bound for~\eqref{eqn:wcos} into an
algorithm; different proof strategies yield different classes of algorithms with
their own pros and cons. 
Deriving the bound is not only important for analyzing the runtime of the
algorithm, but also instrumental in how one thinks about {\em designing} the
algorithm in the first place.
Another significant role that the problem~\eqref{eqn:wcos} plays is in its deep
connection with information theory and (abstract algebraic) group theory.
This paper aims to be a guided tour of these connections.

The notion of worst-case optimality has influenced a couple of other lines 
of inquiries, in parallel query processing~\cite{Ketsman:2017:WOM:3034786.3034788,
DBLP:conf/icdt/KoutrisBS16}, and join processing in the IO 
model~\cite{Hu:2016:TWI:2902251.2902292}.
Furthermore, more than just paperware, $\wcoj$ algorithms have found their way to
academic data management and analytic systems~\cite{DBLP:conf/sigmod/ChuBS15,
DBLP:journals/tods/AbergerLTNOR17,
Kankanamge:2017:GAG:3035918.3056445,
DBLP:journals/pvldb/AmmarMSJ18,
Olteanu:2015:SBF:2751312.2656335}, and are part of two commercial data analytic
engines at {\sf LogicBlox}~\cite{DBLP:conf/sigmod/ArefCGKOPVW15} 
and {\sf RelationalAI}.

%The SIGMOD Records survey~\cite{skew} presented our early understanding of
%bounds and $\wcoj$ algorithms, where only cardinality constraints were
%taken into account. Our understanding of both the problem formulation and solution 
%space has vastly evolved since then. In particular, the problem setting is much
%more general, with $\fd$ and degree constraints taken into account. 
%Connections to information theoretic inequalities and group theory are now
%established.

The author is deeply indebted to Mahmoud Abo Khamis and Dan Suciu, whose insights, 
enthusiasm, and collaborative effort
(both on~\cite{csma,panda} and off official publication records)
in the past few years have helped form the skeleton of the
story that this article is attempting to tell.
The technical landscape has evolved drastically from an early exposition on the
topic~\cite{skew}. 

\subsection{A brief history of bounds and algorithms}
\label{subsec:history}

We start our history tour with the bound~\eqref{eqn:wcos} in the simple setting when all
constraints in $\dc$ are cardinality constraints. Consider, for example, the
following ``triangle query'':
\begin{equation}
   Q_{\triangle}(A,B,C) \leftarrow R(A,B), S(B,C), T(A,C),
   \label{eqn:triangle:query}
\end{equation}
While simple, this is not a toy query.
In social network analysis, counting and enumerating the number of triangles in
a large graph $G=(V,E)$ is an important problem, which corresponds
to~\eqref{eqn:triangle:query} with $R=S=T=E$.
There is a large literature on trying to speed up this one query; see,
e.g.~\cite{Tsourakakis:2009:DCT:1557019.1557111,DBLP:conf/www/SuriV11,Becchetti:2010:EAL:1839490.1839494} and references thereof.

One way to think about the output size bound is to think of $Q_\triangle$ as
containing points $(a,b,c)$ in a three-dimensional space, whose projection onto
the $(A,B)$-plane is contained in $R$, onto the $(B,C)$ plane is contained in
$S$, and onto the $(A,C)$-plane is contained in $T$. 
There is a known geometric inequality shown by Loomis and Whitney in
1949~\cite{MR0031538} which addresses a more general problem: bound the volume
of a convex body in space whose shadows on the coordinate hyperplanes have
bounded areas. The triangle query above corresponds to the discrete measure case,
where ``volume'' becomes ``count''. Specializing to the triangle, Loomis-Whitney
states that $|Q_\triangle| \leq \sqrt{|R| \cdot |S| \cdot |T|}$.
Thus, while studied in a completely different context, Loomis-Whitney's
inequality is our earliest known answer to determining~\eqref{eqn:wcos} for a
special class of join queries. In~\cite{nprr,skew}, we referred to these as the
Loomis-Whitney queries: those are queries where every input atom contains all
but one variable.

In a different context, in 1981 Noga Alon~\cite{MR599482} studied the
problem~\eqref{eqn:wcos} in the case where we want to determine the maximum 
number of occurrences of a given subgraph $H$ in a large graph $G$. ($H$ is the query's
body, and $G$ is the database.)
Alon's interest was to determine the asymptotic behavior of this count, but his
formula was also asymptotically tight. In the triangle case, for example, Alon's
bound is $\Theta(N^{3/2})$, the same as that of Loomis-Whitney.
Here, $N$ is the number of edges in $G$. Alon's general bound is
$\Theta(N^{\rho^*(H)})$, where $\rho^*$ denote the ``fractional edge cover number''
of $H$ (see Section~\ref{sec:prelim}). However, his results were not formulated
in this more modern language.

A paper by Chung et al.~\cite{MR859293} on extremal set theory was especially
influential in our story. The paper proved the ``Product Theorem'' which uses
the {\em entropy argument} connecting a count estimation problem to an entropic
inequality. We will see this argument in action in Section~\ref{sec:bounds}.
The Product theorem is proved using what is now known as {\em Shearer's 
lemma}; a clean formulation and a nice proof of this lemma was given by 
Radhakrishnan~\cite{radhakrishnan}. 

In 1995, Bollob\'as and Thomason~\cite{MR1338683} proved a vast generalization
of Loomis-Whitney's result. Their bound, when specialized down to the discrete
measure and our problem, implies what is now known as the $\agm$-bound (see
below and Corollary~\ref{cor:agm}). The equivalence was shown in Ngo et al.~\cite{nprr}.
The key influence of Bollob\'as-Thomason's result to our story was not the
bound, which can be obtained through Shearer's lemma already, but the inductive
proof based on H\"older's inequality. Their inductive proof suggests a natural
recursive algorithm and its analysis, which lead to the algorithms 
in~\cite{nprr, skew}.

Independently, in 1996 Friedgut and Kahn~\cite{MR1639767} generalized Alon's
earlier result from graphs to hypergraphs, showing that the maximum number of
copies of a hypergraph $\calH$ inside another hypergraph $\calG$ is
$\Theta(N^{\rho^*(\calH)})$. Their argument uses the product theorem from Chung
et al.~\cite{MR859293}.
The entropic argument was used in Friedgut's 2004 paper~\cite{MR2104047}
to prove a beautiful inequality, which we shall call {\em Friedgut's inequality}.
In Theorem~\ref{thm:friedgut} we present an essentially equivalent version
of Friedgut's inequality formulated in a more database-friendly way,
and prove it using the inductive argument from Bollob\'as-Thomason. 
Friedgut's inequality not only implies the $\agm$-bound as a
special case, but also can be used in analyzing the backtracking search
algorithm
presented in Section~\ref{sec:algos}.
Theorem~\ref{thm:friedgut} was stated and used in Beame et
al.~\cite{DBLP:conf/pods/BeameKS14} to analyze parallel query processing; 
Friedgut's inequality is starting to take roots in database theory.

Grohe and Marx (\cite{GM06,DBLP:journals/talg/GroheM14}, 2006) were pushing
boundaries on the parameterized complexity of constraint satisfaction problems.
One question they considered was to determine the maximum possible number of
solutions of a sub-problem defined within a bag of a tree decomposition, given
that the input constraints were presented in the listing representation. This is
exactly our problem~\eqref{eqn:wcos} above, and they proved the bound of
$O(N^{\rho^*(Q)})$ using Shearer's lemma, where $\rho^*(Q)$ denote the 
fractional edge cover number of the hypergraph of the query.
They also presented a join-project query plan running in time
$O(N^{\rho^*(Q)+1})$, which is almost worst-case optimal.

Atserias et al. (\cite{AGM}, 2008) applied the same argument to conjunctive
queries showing what is now known as the $\agm$-bound. 
More importantly, they proved that the bound is
asymptotically tight and studied the average-case output size.
The other interesting result from~\cite{AGM} was from the algorithmic side.
They showed that there is class of queries $Q$ for which a join-project plan
evaluates them in time $O(N^3)$ while any join-only plan
requires $\Omega(N^{\Omega(\log k)})$, where $k$ is the query size.
In particular, join-project plans are strictly better than join-only plans.

Continuing with this line of investigation, Ngo et al. (2012,~\cite{nprr})
presented the $\nprr$ algorithm running in time $\tilde O(N^{\rho^*(Q)})$, and
presented a class of queries for which any join-project plan is worse by a
factor of $\Omega(N^{1-1/k})$ where $k$ is the query size. The class of queries
contains the Loomis-Whitney queries.
The $\nprr$ algorithm and its analysis were overly complicated.
Upon learning about this result, Todd Veldhuizen of {\sf LogicBlox} realized
that his {\sf Leapfrog-Triejoin} algorithm ($\lftj$) can also achieve the same
runtime, with a simpler proof. $\lftj$ is the work-horse join algorithm for the
$\LB$'s datalog engine, and was already implemented since about 2009.
Veldhuizen published his findings in 2014~\cite{LFTJ}.
Inspired by $\lftj$ and its simplicity, Ngo et al.~\cite{skew} presented a simple
recursive algorithm called $\gj$ which also has a compact analysis.

The next set of results extend $\dc$ to more than just cardinality constraints.
Gottlob et al.~\cite{GLVV} extended the $\agm$ bound to handle
$\FD$ constraints, using {\em the} entropy argument. They also proved that
the bound is tight if all $\FD$'s are simple $\FD$s.
Abo Khamis et al.~\cite{csma} observed that the same argument generalizes the
bound to general degree constraints, and addressed the corresponding algorithmic 
question. The bound was studied under the $\FD$-closure lattice formalism, where
they showed that the bound is tight if the lattice is a distributive lattice.
This result is a generalization of Gottlob et al.'s result on the tightness of
the bound under simple $\FD$s.
The connection to information theoretic inequalities and the idea of turning an
inequality proof into an algorithm was also developed with the $\csma$ algorithm
in~\cite{csma}. However, the algorithm was also too complicated, whose analysis
was a little twisted at places.

Finally, in~\cite{panda} we developed a new collection of bounds and proved their
tightness and looseness for {\em disjunctive datalog rules}, a generalization of
conjunctive queries. 
It turns out that under general degree constraints there are two natural classes
of bounds for the quantity~\eqref{eqn:wcos}: the {\em entropic bounds} are tight
but we do not know how to compute them, and the relaxed versions called
{\em polymatroid bounds} are computable via a linear program. 
When there is only cardinality constraints, these two bounds collapse into one
(the $\agm$ bound).
We discuss some of these results in
Section~\ref{sec:bounds}. The idea of reasoning about algorithms' runtimes using
Shannon-type inequalities was also developed to a much more mature and more
elegant degree, with an accompanying algorithm called $\panda$. We discuss what
$\panda$ achieves in more details in Section~\ref{sec:algos}.

\subsection{Organization}
The rest of the paper is organized as follows.
Section~\ref{sec:triangle} gently introduces two ways of bounding the output
size and two corresponding algorithms using the triangle query.
Section~\ref{sec:prelim} presents notations, terminology, and a brief background materials on information theory and properties of the entropy
functions.
Section~\ref{sec:bounds} describes two bounds and two methods for proving output
size bounds on a query given degree constraints. 
This section contains some proofs and observations that have not appeared
elsewhere.
Section~\ref{sec:algos} presents two algorithms evolving naturally from the two
bound-proving strategies presented earlier.
Finally, Section~\ref{sec:opm} lists selected open problems arising from the
topics discussed in the paper.

\section{The triangle query}
\label{sec:triangle}

The simplest non-trivial example illustrating the power of $\wcoj$ algorithms is
the triangle query~\eqref{eqn:triangle:query}.
We use this example to illustrate several ideas: the entropy argument,
two ways of proving an output size bound, and how to derive algorithms from
them.
The main objective of this section is to gently illustrate some of the main
reasoning techniques involved in deriving the bound and the algorithm; 
we purposefully do not present the most compact proofs.
At the end of the section, 
we raise natural questions regarding the assumptions made on the bound
and algorithm to motivate the more general problem formulation discussed in the
rest of the paper.

\paragraph*{The bound} Let $\dom(X)$ denote the domain of attribute $X \in
\{A,B,C\}$. Construct a distribution on $\dom(A) \times \dom(B) \times \dom(C)$ 
where a triple $(a,b,c)$ is selected from the output $Q_\triangle$ uniformly.
Let $H$ denote the entropy function of this distribution, namely for any $\bm X
\subseteq \{A,B,C\}$, $H[\bm X]$ denotes the entropy of the marginal distribution
on the variables $\bm X$.
Then, the following hold:
\begin{align*}
   H[A,B,C] &= \log_2 |Q_\triangle|, \quad \text{ (due to uniformity)}\\
   H[A,B] &\leq \log_2 |R|, \\
   H[B,C] &\leq \log_2 |S|, \\
   H[A,C] &\leq \log_2 |T|.
\end{align*}
The first {\em in}equality holds because the support of the marginal
distribution on $\dom(A) \times \dom(B)$ is a subset of $R(A,B)$, and the entropy is
bounded by the $\log_2$ of the support (see Section~\ref{subsec:it}).
The other two inequalities hold for the same reason.
Hence, whenever there are coefficients $\alpha,\beta,\gamma \geq 0$ for which
\begin{equation}
H[A,B,C] \leq \alpha H[A,B] + \beta H[B,C] + \gamma H[A,C]
\label{eqn:shearer:triangle}
\end{equation}
holds for {\em all} entropy functions $H$, we can derive an output size
bound for the triangle query:
\begin{equation}
|Q_\triangle| \leq |R|^\alpha \cdot |S|^\beta \cdot |T|^\gamma.
\label{eqn:triangle:agm:product}
\end{equation}
In Section~\ref{sec:bounds} we will show that~\eqref{eqn:shearer:triangle} holds for all
entropy function $H$ {\em if and only if} $\alpha+\beta\geq 1$, $\beta+\gamma
\geq 1$, and $\alpha+\gamma \geq 1$, for non-negative $\alpha,\beta,\gamma$.
This fact is known as {\em Shearer's inequality}, though Shearer's
original statement is weaker than what was just stated.

One consequence of Shearer's inequality is that, 
to obtain the best possible bound, we will want to minimize the right-hand-side
(RHS) of~\eqref{eqn:triangle:agm:product} subject to the above constraints:
\begin{align}
   \log_2 |Q_\triangle| \leq \min 
   & \quad \alpha \log_2 |R| + \beta \log_2 |S| + \gamma \log_2 |T| 
   \label{eqn:triangle:agm}\\
   \text{s.t.} & \quad \alpha+\beta \geq 1, \\
   & \quad \alpha+\gamma\geq 1,\\
   & \quad \beta+\gamma \geq 1, \\
   & \quad \alpha, \beta, \gamma \geq 0. 
\end{align}
This bound is known as the {\em $\agm$-bound} for $Q_\triangle$.
It is a direct consequence of Friedgut's inequality (Theorem~\ref{thm:friedgut}).
%$\agm$~\cite{AGM} also showed that it is assymptotically tight, in the sense that there is
%an input instance whose output size essentially matches the bound.

\paragraph*{Algorithms} Let $N = \max\{|R|,|S|,|T|\}$, and $(\alpha^*,\beta^*,\gamma^*)$ denote
an optimal solution to the LP~\eqref{eqn:triangle:agm} above. A $\wcoj$ algorithm 
needs to be able to answer $Q_\triangle$ in time $\tilde
O(N+|R|^{\alpha^*}|S|^{\beta^*}|T|^{\gamma^*})$, where $\tilde O$ hides a {\em single} $\log
N$ factor. The feasible region of~\eqref{eqn:triangle:agm} is a $3$-dimensional
simplex. Without loss of generality, we can assume that $(\alpha^*,\beta^*,\gamma^*)$ is
one of the $4$ vertices of the simplex, which are
$(1,1,0)$, $(1, 0,1)$, $(0,1,1)$, and $(.5,.5,.5)$.
If $(\alpha^*,\beta^*,\gamma^*) = (1,1,0)$, then the traditional join plan
$(R \Join S) \Join T$
has the desired runtime of $\tilde O(|R|\cdot |S|) = 
\tilde O(|R|^{\alpha^*}|S|^{\beta^*}|T|^{\gamma^*})$, modulo $\tilde O(N)$ preprocessing
time. 

Consequently, the only interesting case is when
$(\alpha^*,\beta^*,\gamma^*) = (.5,.5,.5)$.
It is easy to see that this is optimal to LP~\eqref{eqn:triangle:agm} when
the product of sizes of any two relations from $R$, $S$, and $T$ is greater than
the size of the third relation.
To design an algorithm running in 
$\tilde O(N+\sqrt{|R|\cdot |S|\cdot |T|})$-time, 
we draw inspiration from two different {\em proofs} of
the bound~\eqref{eqn:triangle:agm}.

{\bf First Algorithm.} Write $\bm 1_E$ to denote the indicator variable for
    the Boolean event $E$; for example $\bm 1_{R(a,b)}$ is $1$ if $(a,b) \in R$
    and $0$ otherwise.
    Let $\sigma$ denote the relational selection operator.
    The Bollob\'as-Thomason's argument for
    proving~\eqref{eqn:triangle:agm} goes as follows.
    \begin{align}
       |Q_\triangle| 
       &= \sum_a \sum_b \sum_c \bm 1_{R(a,b)}\bm 1_{S(b,c)}\bm 1_{T(a,c)}\\
       &= \sum_a \sum_b \bm 1_{R(a,b)} \sum_c \bm 1_{S(b,c)} \bm 1_{T(a,c)}\\
       &\leq \sum_a \sum_b \bm 1_{R(a,b)} \sqrt{\sum_c \bm 1_{S(b,c)}} \cdot
       \sqrt{\sum_c \bm 1_{T(a,c)}}\\
       &= \sum_a \sum_b \bm 1_{R(a,b)} 
       \sqrt{|\sigma_{B=b}S|}
       \cdot \sqrt{|\sigma_{A=a}T|} \label{eqn:13}\\
       &= \sum_a \sqrt{|\sigma_{A=a}T|}
       \sum_b \bm 1_{R(a,b)} \cdot
       \sqrt{|\sigma_{B=b}S|}\\
       &\leq \sum_a \sqrt{|\sigma_{A=a}T|}
       \sqrt{\sum_b \bm 1_{R(a,b)}} \cdot
       \sqrt{\sum_b |\sigma_{B=b}S|}\\
       &= \sum_a \sqrt{|\sigma_{A=a}T|}
       \sqrt{|\sigma_{A=a}R|}
       \sqrt{|S|}\\
       &= 
       \sqrt{|S|} \cdot
       \sum_a \sqrt{|\sigma_{A=a}T|}
       \sqrt{|\sigma_{A=a}R|}\\
       &\leq \sqrt{|R| \cdot |S| \cdot |T|}.
    \end{align}
    All three inequalities follow from Cauchy-Schwarz. Tracing the
    inequalities back to $|Q_\triangle|$, Algorithm~\ref{algo:1} emerges. 
    \begin{algorithm}[t]
       \For {$a \in \pi_AR \cap \pi_A T$} {
          \For {$b \in \pi_B \sigma_{A=a} R \cap \pi_B S$} {
             \For {$c \in \pi_C \sigma_{B=b} S \cap \pi_C \sigma_{A=a} T$} {
                Report $(a,b,c)$\;
            }
         }
       }
       \caption{based on H\"older's inequality proof}
       \label{algo:1}
    \end{algorithm}
    The analysis is based on only a single assumption, that we can loop through
    the intersection of two sets $X$ and $Y$ in time bounded by $\tilde
    O(\min\{|X|,|Y|\})$.
    This property can be satisfied with sort-merge or simple hash join when we
    iterate through the smaller of the two sets and look up in the hash table of
    the other.
    For a fixed binding $(a,b)$, the inner-most loop runs in time
       \[ \min\left\{
          |\pi_C \sigma_{B=b} S|, |\pi_C \sigma_{A=a} T|
          \right\}
       \leq \sqrt{|\sigma_{B=b} S| \cdot |\sigma_{A=a} T|}. \]
    A binding $(a,b)$ gets in the inner loop only if $(a,b) \in R$, and so the
    total amount of work is
    \begin{equation}
       \tilde O\left(\sum_a \sum_b \bm 1_{R(a,b)}\sqrt{|\sigma_{B=b} S| \cdot |\sigma_{A=a}
    T|}\right).
    \end{equation}
    Compare this with~\eqref{eqn:13}, 
    and the runtime analysis is completed.

{\bf Second Algorithm.} This algorithm is inspired by a proof of
    inequality~\eqref{eqn:shearer:triangle}, which
    implies~\eqref{eqn:triangle:agm}.
    In this particular case~\eqref{eqn:shearer:triangle} can be written as
\begin{equation}
2H[A,B,C] \leq 
H[A,B]+H[B,C]+H[A,C].
\label{eqn:triangle:2}
\end{equation}
Using the chain rule (eq.~\eqref{eqn:entropy:chain:rule}) and the 
submodularity rule (eq.~\eqref{eqn:submodularity}) for entropic functions,
the inequality can be proved as follows.
\begin{align}
H[A,B]+H[B,C]+H[A,C]
&= H[A] + H[B \suchthat A] + H[B,C]+H[A,C] \label{eqn:entropic:chain:rule}\\
&= (H[A] + H[B,C]) + (H[B \suchthat A]+H[A,C])\\
&\geq (H[A \suchthat B,C] + H[B,C]) + (H[B \suchthat A,C]+H[A,C])\\
&= 2 H[A,B,C].
\end{align}
The first replacement $H[A,B] \to H[A] + H[B \suchthat A]$ is interpreted as a
{\em decomposition} of the relation $R(A,B)$ into two parts ``heavy'' and
``light''. After applying submodularity, two {\em compositions} are performed
to obtain two copies of $H[A,B,C]$: 
$H[A \suchthat B,C] + H[B,C] \to H[A,B,C]$ and
$H[B \suchthat A,C]+H[A,C] \to H[A,B,C]$. These correspond to join operators.
Algorithm~\ref{algo:2} has the pseudo-code.
It is remarkable how closely the algorithm mimics the entropy proof.
    \begin{algorithm}[t]
       $\theta \leftarrow \sqrt{\frac{|R|\cdot |S|}{|T|}}$\;
       $R^{\heavy} \leftarrow \{(a,b) \in R \ : \ |\sigma_{A=a}R| > \theta\}$\;
       $R^{\light} \leftarrow \{(a,b) \in R \ : \ |\sigma_{A=a}R| \leq \theta\}$\;
       \Return  $[(R^{\heavy} \Join S) \ltimes T] \ \cup \ [(R^\light \Join T)
       \ltimes S]$\;
       \caption{based on entropy inequality proof}
       \label{algo:2}
    \end{algorithm}

The analysis is also compact. Note that $|R^\heavy| \leq |R|/\theta$ and thus
       \[ |R^{\heavy} \Join S| \leq \frac{|R|\cdot |S|}{\theta} =
       \sqrt{|R|\cdot|S|\cdot|T|}. \]
       In the other case,
       $|R^{\light} \Join T| \leq |T| \cdot \theta =
       \sqrt{|R|\cdot|S|\cdot|T|}.$
       This completes the analysis.

\paragraph*{Follow-up questions} 
In a more realistic setting, we know more about the input than just the
cardinalities. In a database there may (and will) be $\FD$s. In a graph we may
know the maximum degree of a vertex. How do the bounds and algorithms change
when we take such information into account? Which of the above two bounds and
algorithms generalize better in the vastly more general setting? We explore
these questions in the remainder of this paper.

\section{Preliminaries}
\label{sec:prelim}

Throughout the paper, we use the following convention.  The                     
non-negative reals, rationals, and integers are denoted by                      
$\R_+,\Q_+$, and $\N$ respectively. For a positive integer $n$, $[n]$ denotes   
the set $\{1,\dots,n\}$.                                                        
                                                                                
Functions $\log$ without a base specified are                                   
base-$2$, i.e. $\log = \log_2$. Uppercase $A_i$ denotes a                       
variable/attribute, and lowercase $a_i$ denotes a value in the                  
discrete domain $\dom(A_i)$ of the variable.  For any subset                    
$S\subseteq [n]$, define $\mv A_S = (A_i)_{i\in S}$,                            
$\mv a_S = (a_i)_{i\in S} \in \prod_{i\in S}\dom(A_i)$.  In                     
particular, $\mv A_S$ is a tuple of variables and $\mv a_S$ is a tuple          
of specific values with support $S$.  
We also use $\bm X_S$ to denote variables and $\bm x_S$, $\bm t_S$ to denote 
value tuples in the same way.

\subsection{Queries and degree constraints}
\label{subsec:dc}

A {\em multi-hypergraph} is a hypergraph where edges may occur more than once.
We associate a full conjunctive query $Q$ to a multi-hypergraph                       
$\calH := ([n],\calE)$, $\calE \subseteq 2^{[n]}$; the query is written as
\begin{equation}                                                                
   Q(\bm A_{[n]}) \leftarrow \bigwedge_{F\in\calE} R_F(\bm A_F), \label{eqn:conjunctive:query}
\end{equation}                                                                  
with variables $A_i$, $i \in [n]$, and atoms
$R_F$, $F \in \calE$. 

\bdefn[Degree constraint]
A {\em degree constraint} is a triple $(X, Y, N_{Y|X})$, where 
$X \subsetneq Y \subseteq [n]$  and $N_{Y|X} \in \N$.
The relation $R_F$ is said to {\em guard} the degree constraint $(X,Y,N_{Y|X})$
if $Y\subseteq F$ and 
\begin{equation}                                                                
   \deg_F(\bm A_Y | \bm A_X)  := \max_{\bm t} |\Pi_{\bm A_Y}(\sigma_{\bm     
   A_X=\bm t}(R_F))| \leq N_{Y|X}.
\end{equation}                                                                  
Note that a given relation may guard multiple degree constraints.
Let $\dc$ denote a set of degree constraints. The input database $\bm D$ 
is said to {\em
satisfy} $\dc$ if every constraint in $\dc$ has a guard, in which case we write
$\bm D \models \dc$.
\edefn

A {\em cardinality constraint} is an                                            
assertion of the form $|R_F| \leq N_F$, for some $F \in \calE$; it is exactly   
the degree constraint $(\emptyset, F, N_{F|\emptyset})$ guarded by $R_F$.
A {\em functional dependency} $\bm A_X \rightarrow \bm A_Y$ is a                
degree constraint with $N_{X\cup Y|X}=1$.                                       
In particular, degree constraints strictly generalize both cardinality          
constraints and FDs.                                                            

Our problem setting is general, where we are given a query of the
form~\eqref{eqn:conjunctive:query} and a set $\dc$ of degree constraints
satisfied by the input database $\bm D$.
The first task is to find a good upper bound, or determine exactly the
quantity $\sup_{\bm D \models \dc} |Q(\bm D)|$, the worst-case output size
of the query given that the input satisfies the degree constraints.
The second task is to design an algorithm running in time as close to the bound
as possible. 

Given a multi-hypergraph $\calH = ([n], \calE)$, define its corresponding
``fractional edge cover polytope'':
\begin{equation*}
   \fecp(\calH) := \left\{
      \bm \delta = (\delta_F)_{F\in\calE} \suchthat \bm \delta \geq \bm 0 \wedge
      \sum_{F : v \in F} \delta_F \geq 1, \forall v \in [n]
   \right\}.
\end{equation*}
Every point $\bm\delta  \in \fecp(\calH)$ is called a 
{\em fractional edge cover} of $\calH$.
The quantity 
\[ \rho^*(\calH) := \min \left\{ \sum_{F\in\calE}\delta_F \suchthat \bm\delta \in
   \fecp(\calH) \right\} \]
is called the {\em fractional edge cover number} of $\calH$.

\subsection{Information theory}
\label{subsec:it}

The books~\cite{MR2239987,Yeung:2008:ITN:1457455} are good references on information theory. We
extract only simple facts needed for this paper.
Consider a joint probability distribution $\calD$ on $n$ discrete
variables $\bm A = (A_i)_{i\in [n]}$ and a probability mass function $\pr$.
The {\em entropy function} associated with $\calD$ 
is a function $H : 2^{\bm A} \to \R_+$, where 
\begin{equation}
   H[\bm A_F] := \sum_{\bm a_F \in \prod_{i\in F} \dom(A_i)} \pr[\bm A_F =\bm
   a_F] \log \frac{1}{\pr[\bm A_F = \bm a_F]}
\end{equation}
is the entropy of the marginal distribution on $\bm A_F$.
To simplify notations, we will also write $H[F]$ for $H[\bm A_F]$, turning $H$
into a set function $H : 2^{[n]} \to \R_+$.
For any $F \subseteq [n]$, define the ``support'' of the marginal distribution
on $\bm A_F$ to be
\begin{equation}
   \supp_F(\calD) := \left\{ \bm x_F \in \prod_{i\in F}\dom(A_i) \suchthat
   \pr[\bm A_F = \bm x_F] > 0 \right\}.
\end{equation}
Given $X \subseteq Y \subseteq [n]$, define the {\em conditional
entropy} to be
\begin{equation}
   H[Y \suchthat X] := H[Y] - H[X].
   \label{eqn:entropy:chain:rule}
\end{equation}
This is also known as the {\em chain rule} of entropy.
The following facts are basic and fundamental in information theory:
\begin{align}
   H[\emptyset] &= 0\\
   H[X] & \leq \log |\supp_X(\calD)| && \forall X\subseteq [n]
   \label{eqn:jensen}\\
   H[X] &\leq H[Y] && \forall X \subseteq Y \subseteq [n]
   \label{eqn:monotonicity}\\
   H[X \cup Y \suchthat Y] &\leq H[X \suchthat X \cap Y] && \forall X,Y \subseteq [n]
   \label{eqn:submodularity}
\end{align}
Inequality~\eqref{eqn:jensen} follows from Jensen's inequality and the concavity
of the entropy function. Equality holds if and only if the marginal distribution
on $X$ is {\em uniform}. 
Entropy measures the ``amount of uncertainty'' we have: the more uniform the
distribution, the less certain we are about where a random point
is in the space.
Inequality~\eqref{eqn:monotonicity} is the {\em monotonicity} property: adding
more variables {\em increases} uncertainty.
Inequality~\eqref{eqn:submodularity} is the {\em submodularity} property:
conditioning on more variables {\em reduces} uncertainty.\footnote{$H[X
\suchthat X \cap Y] \geq H[X \suchthat (X\cap Y) \cup (Y\setminus X)]
= H[X \suchthat Y] = H[X \cup Y \suchthat Y]$.}

A function $f : 2^{\calV} \to \R_+$ is called a (non-negative)                  
{\em set function} on $\calV$.                                                  
A set function $f$ on $\calV$ is {\em modular} if                               
$f(S) = \sum_{v\in S} f(\{v\})$ for all $S\subseteq \calV$,                     
is {\em monotone} if $f(X) \leq f(Y)$ whenever $X \subseteq Y$,                 
is {\em subadditive} if $f(X\cup Y)\leq f(X)+f(Y)$                              
for all $X,Y\subseteq \calV$,                                                   
and is {\em submodular} if $f(X\cup Y)+f(X\cap Y)\leq f(X)+f(Y)$                
for all $X,Y\subseteq \calV$.                                                   
Let $n$ be a positive integer.                                                  
A function $h : 2^{[n]} \to \R_+$ is said to be {\em entropic} if there 
is a joint distribution on $\mv A_{[n]}$ with entropy function $H$ 
such that $h(S) = H[S]$ for all $S\subseteq [n]$.
We will write $h(S)$ and $h(\mv A_S)$ interchangeably, depending on context.

Unless specified otherwise,                                                     
we will only consider {\em non-negative} and {\em monotone} set functions       
$f$ for which $f(\emptyset) = 0$; this assumption will be implicit in the       
entire paper.                                                                   
                                                                                
\bdefn%[Commonly used classes of set functions]                                 
Let $\Mod_n$, $\sa_n$, and $\Gamma_n$ denote the set of all (non-negative and   
monotone) modular, subadditive, and submodular set functions on                 
$\calV$, respectively.                                                          
Let $\Gamma^*_n$ denote the set of all entropic functions on $n$ variables, and 
$\overline\Gamma^*_n$ denote its topological closure.                           
The set $\Gamma_n$ is called the set of {\em polymatroidal functions}, or simply
{\em polymatroids}.
\edefn                                                                          

The notations $\Gamma_n, \Gamma^*_n,\overline \Gamma^*_n$ are        
standard in information theory.
It is known~\cite{Yeung:2008:ITN:1457455} that $\Gamma^*_n$ is a cone
which is not topologically closed. And hence, when optimizing over this cone we
take its topological closure $\overline\Gamma^*_n$, which is convex.
The sets $\Mod_n$ and $\Gamma_n$ are clearly polyhedral cones.

As mentioned above, entropic functions satisfy non-negativity, monotonicity, and
submodularity. Linear inequalities regarding entropic functions derived from
these three properties are called {\em Shannon-type} inequalities.
For a very long time, it
was widely believed that Shannon-type inequalities form a complete set of
linear inequalities satisfied by entropic functions, namely $\overline\Gamma^*_n
= \Gamma_n$. This indeed holds for $n \leq 3$, for example.
However, in 1998, in a breakthrough paper in information theory,
Zhang and Yeung~\cite{DBLP:journals/tit/ZhangY98} presented 
a new inequality which cannot be implied by Shannon-type
inequalities. Their result proved that, $\overline\Gamma^*_n \subsetneq
\Gamma_n$ for any $n \geq 4$. 
Lastly the following chain of inclusion is known~\cite{Yeung:2008:ITN:1457455}
\begin{equation}
   \Mod_n \subseteq \Gamma^*_n \subseteq \overline\Gamma^*_n \subseteq \Gamma_n
   \subseteq \sa_n.
   \label{eqn:inclusion:chain}
\end{equation}
When $n \geq 4$, all of the containments are {\em strict}.
                                                                                
%\begin{figure}[t!]                                                             
%\centering \input{setFunc}                                                      
%   \caption{Hierarchy of set functions}                                         
%\label{fig:set:functions}                                                       
%\end{figure} 

%\input{bounds}

\section{Output size bounds}
\label{sec:bounds}

This section addresses the following question: given a query $Q$ and a set
of degree constraints $\dc$, determine $\sup_{\bm D\models \dc}|Q(\bm D)|$ or at
least a good upper bound of it.

\subsection{Cardinality constraints only}
\label{eqn:cardinality:bounds}

Friedgut's inequality is essentially equivalent to H\"older's inequality. 
Following Beame et al.~\cite{DBLP:conf/pods/BeameKS14}, who used the inequality 
to analyze parallel query processing algorithms,
we present here a version that is more database-friendly.
We also present a proof of Friedgut's inequality using H\"older's inequality, 
applying the same induction strategy used in the proof of Bollob\'as-Thomason's 
inequality~\cite{MR1338683} and the ``query decomposition lemma'' in~\cite{skew}.

\bthm[Friedgut~\cite{MR2104047}]
Let $Q$ denote a full conjunctive query with (multi-) hypergraph $\calH = ([n], \calE)$
and input relations $R_F$, $F \in \calE$. Let $\bm\delta =
(\delta_F)_{F\in\calE}$ denote a fractional edge cover of $\calH$.
For each $F\in \calE$, let $w_F : \prod_{i \in F}\dom(A_i) \to \R_+$ denote an
arbitrary
non-negative weight function. Then, the following holds
\begin{equation}
   \sum_{\bm a \in Q} \prod_{F\in \calE} [w_F(\bm a_F)]^{\delta_F}
   \leq
   \prod_{F\in\calE}
   \left(
   \sum_{\bm t \in R_F} w_F(\bm t)
   \right)^{\delta_F}
   \label{eqn:friedgut:ineq}
\end{equation}
\label{thm:friedgut}
\ethm
%\bp
\begin{proof}
We induct on $n$. When $n=1$, the inequality is exactly generalized H\"older
inequality~\cite{MR89d:26016}. Suppose $n>1$,
and -- for induction purposes -- define a new query $Q'$ whose hypergraph is
$\calH'=([n-1],\calE')$, new fractional edge cover $\bm\delta' =
(\delta'_F)_{F\in \calE'}$ for $\calH'$, and new weight functions $w'_F$ for each $F \in \calE'$
as follows:
\begin{align}
   \partial(n) &:= \{F \in \calE\suchthat n \in F\}, \\
   \calE' &:= \{ F \suchthat F \neq \emptyset \wedge (F \in \calE \setminus \partial(n)
   \vee F \cup \{n\} \in \calE) \} \\
   R'_F &:= \begin{cases}
     R_F & F \in \calE \setminus \partial(n) \\
      \pi_{\bm A_F} R_{F  \cup \{n\}} & F \cup \{n\} \in \calE
   \end{cases}
      && F \in \calE'\\
   \delta'_F &:= \begin{cases}
     \delta_F & F \in \calE \setminus \partial(n) \\
      \delta_{F \cup \{n\}} & F \cup \{n\} \in \calE
   \end{cases}
      && F \in \calE'\\
   Q' &:= \ \Join_{F \in \calE'} R'_F \\
   w'_F(\bm a_F) &:= \begin{cases}
      w_F(\bm a_F) & F \in \calE - \partial(n)\\
      \sum_{a_n} w_{F\cup \{n\}}(\bm a_{F\cup\{n\}})\bm 1_{R_{F \cup\{n\}}(\bm a_{F
      \cup\{n\}})} & F\cup \{n\} \in \calE
   \end{cases} && F \in \calE'
\end{align}
Then, by noting that the tuple
$\bm a = (a_1,\dots,a_n)\in \prod_{i=1}^n \dom(A_i)$
belongs to $Q$ if and only if 
   $\prod_{F \in\calE} \bm 1_{R_F(\bm a_F)} = 1$, we have\footnote{We use the
   convention that $0^0=0$.}
\begin{align*}
   & \sum_{\bm a \in Q} \prod_{F\in \calE} [w_F(\bm a_F)]^{\delta_F} 
   = \sum_{\bm a_{[n-1]}} \sum_{a_n} 
      \prod_{F\in \calE} [w_F(\bm a_F)\bm 1_{R_F(\bm A_F)}]^{\delta_F}\\
   &= \sum_{\bm a_{[n-1]}} 
      \prod_{F\in \calE\setminus\partial(n)} [w_F(\bm a_F)\bm 1_{R_F(\bm a_F)}]^{\delta_F}
      \sum_{a_n} 
      \prod_{F\in \calE_n} [w_F(\bm a_F)\bm 1_{R_F(\bm a_F)}]^{\delta_F}\\
   &\leq \sum_{\bm a_{[n-1]}} 
      \prod_{F\in \calE\setminus\partial(n)} [w_F(\bm a_F)\bm 1_{R_F(\bm a_F)}]^{\delta_F}
      \prod_{F\in \calE_n} 
      \left[ \sum_{a_n} w_F(\bm a_F)\bm 1_{R_F(\bm a_F)} 
      \right]^{\delta_F}\\
   &= \sum_{\bm a_{[n-1]}} 
      \prod_{F\in \calE\setminus\partial(n)} [w_F(\bm a_F)\bm 1_{R_F(\bm a_F)}]^{\delta_F}
      \prod_{\substack{F\in \calE' \\ F\cup \{n\} \in \calE}} \left[ \sum_{a_n}
      w_{F\cup\{n\}}(\bm a_{F\cup\{n\}})\bm 1_{R_{F\cup\{n\}}(\bm a_{F\cup\{n\}})} \right]^{\delta_{F\cup\{n\}}}
      \prod_{\substack{F \in \calE\\ F = \{n\}}} \left[ \sum_{\bm t \in R_F}
      w_F(\bm t) \right]^{\delta_F}
      \\
   &= 
      \prod_{\substack{F \in \calE\\ F = \{n\}}} \left[ \sum_{\bm t \in R_F} w_F(\bm t) \right]^{\delta_F}
      \cdot
   \sum_{\bm a_{[n-1]}} 
      \prod_{F\in \calE\setminus\partial(n)} [w'_F(\bm a_F)\bm 1_{R'_F(\bm a_F)}]^{\delta'_F}
      \prod_{\substack{F\in \calE' \\ F\cup \{n\} \in \calE}} \left[ 
      w'_{F}(\bm a_{F})\bm 1_{R'_{F}(\bm a_{F})} \right]^{\delta'_{F}}
      \\
   &= 
      \prod_{\substack{F \in \calE\\ F = \{n\}}} \left[ \sum_{\bm t \in R_F} w_F(\bm t) \right]^{\delta_F}
      \cdot
      \sum_{\bm a_{[n-1]} \in Q'} \prod_{F\in \calE'} [w'_F(\bm a_F)]^{\delta'_F}
      \\
   &\leq 
      \prod_{\substack{F \in \calE\\ F = \{n\}}} \left[ \sum_{\bm t \in R_F} w_F(\bm t) \right]^{\delta_F}
      \cdot
      \prod_{F\in \calE'} 
      \left[
      \sum_{\bm t \in R'_F}
      w'_F(\bm t) \right]^{\delta'_F}
      \\
   &= 
      \prod_{\substack{F \in \calE\\ F = \{n\}}} \left[ \sum_{\bm t \in R_F} w_F(\bm t) \right]^{\delta_F}
      \cdot
      \prod_{F\in \calE \setminus \partial(n)} \left[ \sum_{\bm t \in R'_F} w'_F(\bm t) \right]^{\delta'_F}
      \prod_{\substack{F\in \calE' \\ F \cup \{n\} \in \calE}} \left[ \sum_{\bm t \in R'_F} w'_F(\bm t) \right]^{\delta'_F}
      \\
   &= 
      \prod_{\substack{F \in \calE\\ F = \{n\}}} \left[ \sum_{\bm t \in R_F} w_F(\bm t) \right]^{\delta_F}
      \cdot
      \prod_{F\in \calE \setminus \partial(n)} \left[ \sum_{\bm t \in R_F} w_F(\bm t) \right]^{\delta_F}
      \prod_{\substack{F\in \calE' \\ F \cup \{n\} \in \calE}} \left[ \sum_{\bm
      t \in R'_F} \sum_{a_n} w_{F \cup \{n\}} (\bm t,a_n) \bm 1_{R_{F \cup
      \{n\}}(\bm t,a_n)}
      \right]^{\delta_{F\cup\{n\}}}
      \\
   &=
   \prod_{F\in \calE} \left( \sum_{\bm t \in R_F} w_F(\bm t)\right)^{\delta_F}
\end{align*}
The first inequality follows from H\"older's inequality
and the fact that $\bm \delta$ is a fractional edge cover; in particular, $\sum_{F \in \calE_n}
\delta_F \geq 1$. The second inequality is the induction hypothesis.
\ep
By setting all weight functions to be identically $1$, we obtain 
\bcor[$\agm$-bound~\cite{AGM}]
Given the same setting as that of Theorem~\ref{thm:friedgut}, we have
\begin{equation}
  |Q| \leq \prod_{F \in \calE} |R_F|^{\delta_F}.  \label{eqn:agm}
\end{equation}
   In particular, let $N = \max_{F\in\calE} |R_F|$ then $|Q| \leq
   N^{\rho^*(\calH)}$.
\label{cor:agm}
\ecor

\subsection{General degree constraints}

To obtain a bound in the general case, we employ the entropy argument, 
which by now is widely
used in extremal combinatorics~\cite{MR2865719,MR859293,radhakrishnan}. 
In fact, Friedgut~\cite{MR2104047} proved Theorem~\ref{thm:friedgut} using an
entropy argument too.
The particular argument below
can be found in the first paper mentioning Shearer's inequality~\cite{MR859293}, 
and a line of follow-up work~\cite{MR1639767,radhakrishnan,GLVV,csma,panda}.

Let $\bm D \models \dc$ be any database instance satisfying the input degree 
constraints. Construct a distribution $\calD$ on $\prod_{i\in [n]}\dom(A_i)$ by
picking uniformly a tuple $\bm a_{[n]}$ from the output $Q(\bm D)$.
Let $H$ denote the corresponding entropy function.
Then, due to uniformity we have $\log_2|Q(\bm D)| = H([n])$.
Now, consider any degree constraint $(X,Y,N_{Y|X}) \in \dc$ guarded by an input
relation $R_F$. From~\eqref{eqn:jensen} it follows that $H[Y \suchthat X] \leq
\log N_{Y|X}$. Define the collection $\hdc$ 
of set functions satisfying the degree
constraints $\dc$:
\begin{equation*}
   \hdc := \{ h \suchthat 
   h(Y)-h(X) \leq \log N_{Y|X}, \forall (X,Y,N_{Y|X})\in \dc\}.
\end{equation*}
Then, the entropy argument immediately gives the following result, first
explicitly formulated in~\cite{panda}:
\bthm[From~\cite{csma,panda}]
Let $Q$ be a conjunctive query and $\dc$ be a given set of degree constraints,
then for any database $\bm D$ satisfying $\dc$, we have
\begin{align}
   \log |Q(\bm D)| &\leq
   \max_{h \in \overline\Gamma^*_n \cap \hdc} h([n]) && \text{(entropic bound)}\\
   &\leq
   \max_{h \in \Gamma_n \cap \hdc} h([n]) && \text{(polymatroid bound)}
   \label{eqn:polymatroid:bound}
\end{align}
   Furthermore, the entropic bound is asymptotically tight and the polymatroid
   bound is not.
\ethm
The polymatroid relaxation follows from the chain of
inclusion~\eqref{eqn:inclusion:chain}; the relaxation is necessary because we do not know
how to compute the entropic bound.
Also from the chain of inclusion, we remark that while the set $\sa_n$ is not 
relevant to our story, we can further move from $\Gamma_n$ to $\sa_n$ and end 
up with the {\em integral} edge cover number~\cite{panda}.

Table~\ref{tab:bounds}, extracted from~\cite{panda}, summarizes our current
state of knowledge on the tightness and looseness of these two bounds.
\colorlet{clr:entropy}{blue}
\colorlet{clr:polymatroid}{red}
\begin{table}[th!]
   \begin{tabular}{|c|c|c|}
      \hline
      \rowcolor{gray!30}
      Bound &
      {\color{clr:entropy}Entropic} Bound&
      {\color{clr:polymatroid}Polymatroid} Bound\\
      \hline\hline
      %%%%%%%%%%%%%%%%%%%%%%%%%%%%%%%%%%%%%%%%%%%%%%%%%%%%%%%%%%%%%%%%%%%%
      \begin{tabular}{c}
         \\
         Definition\\
         ~
      \end{tabular}&
      \begin{tabular}{c}
         $\displaystyle{\sup_{\bm D \models \dc} \log|Q(\bm D)| \leq \max_{h \in
         {\color{clr:entropy}\overline\Gamma^*_n} \cap \hdc}h([n])}$\\
         (See~\cite{GLVV,csma})
      \end{tabular}&
      \begin{tabular}{c}
         $\displaystyle{\sup_{\bm D \models \dc} \log|Q(\bm D)| \leq\max_{h \in {\color{clr:polymatroid}\Gamma_n} \cap \hdc}h([n])}$\\
      (See~\cite{GLVV,csma})
      \end{tabular}\\
      \cline{1-3}
      %%%%%%%%%%%%%%%%%%%%%%%%%%%%%%%%%%%%%%%%%%%%%%%%%%%%%%%%%%%%%%%%%%%%
      \begin{tabular}{c}
         $\dc$ contains only\\
         cardinality constraints 
      \end{tabular}&
      \begin{tabular}{c}
         $\agm$ bound~\cite{DBLP:journals/talg/GroheM14,AGM}\\
         (Tight~\cite{AGM})
      \end{tabular}&
      \begin{tabular}{c}
         $\agm$ bound~\cite{DBLP:journals/talg/GroheM14,AGM}\\
         (Tight~\cite{AGM})
      \end{tabular}\\
      \cline{1-3}
      %%%%%%%%%%%%%%%%%%%%%%%%%%%%%%%%%%%%%%%%%%%%%%%%%%%%%%%%%%%%%%%%%%%%
      %$\cc$ and $\FD$&
      \begin{tabular}{c}
      $\dc$ contains only \\
      cardinality and 
      FD constraints 
      \end{tabular}&
      \begin{tabular}{c}
         {\color{clr:entropy}Entropic} Bound for $\FD$~\cite{GLVV}\\
         ({\color{clr:entropy}Tight}~\cite{szymon-2015})
      \end{tabular}&
      \begin{tabular}{c}
         {\color{clr:polymatroid}Polymatroid} Bound for $\FD$~\cite{GLVV}\\
         {({\color{clr:polymatroid}Not tight} \cite{panda})}
      \end{tabular}\\
      \cline{1-3}
      %%%%%%%%%%%%%%%%%%%%%%%%%%%%%%%%%%%%%%%%%%%%%%%%%%%%%%%%%%%%%%%%%%%%
      \begin{tabular}{c}
         $\dc$ is a general \\
         set of degree constraints %($\dc$)
      \end{tabular}&
      \begin{tabular}{c}
         {\color{clr:entropy}Entropic} Bound for $\dc$~\cite{csma}\\
         {({\color{clr:entropy}Tight} \cite{panda})}
      \end{tabular}&
      \begin{tabular}{c}
         {\color{clr:polymatroid}Polymatroid} Bound for $\dc$~\cite{csma}\\
         {({\color{clr:polymatroid}Not tight} \cite{panda})}
      \end{tabular}\\
      \hline\hline
      %%%%%%%%%%%%%%%%%%%%%%%%%%%%%%%%%%%%%%%%%%%%%%%%%%%%%%%%%%%%%%%%%%%%
   \end{tabular}
\caption{Summary of entropic and polymatroid size bounds for full conjunctive queries along with their tightness properties.}
\label{tab:bounds}
\end{table}
The entropic bound is asymptotically tight, i.e. there are arbitrarily large
databases $\bm D \models \dc$ for which $\log |Q(\bm D)|$ approaches the entropic
bound. The polymatroid bound is not tight, i.e. there exist a query and degree
constraints for which its distance from the entropic bound
is arbitrarily large.

The tightness of the entropic bound is proved using a very interesting
connection between information theory and group theory first observed in Chan
and Yeung~\cite{DBLP:journals/tit/ChanY02}.
Basically, given any entropic function $h \in \overline\Gamma^*_n \in \hdc$, 
one can construct a database instance $\bm D$ which satisfies all degree
constrains $\dc$ and $\log |Q(\bm D)| \geq h([n])$. The database instance is
constructed from a system of (algebraic) groups derived from the entropic
function.

The looseness of the polymatroid bound follows from Zhang and Yeung 
result~\cite{DBLP:journals/tit/ZhangY98} mentioned in Section~\ref{subsec:it}.
In~\cite{panda}, we exploited Zhang-Yeung non-Shannon-type
inequality and constructed a query for which the optimal polymatroid solution
$h^*$ to
problem~\eqref{eqn:polymatroid:bound} strictly belongs to $\Gamma_n -
\overline\Gamma^*_n$. This particular $h^*$ proves the gap between the two
bounds, which we can then magnify to an arbitrary degree by scaling up the
degree constraints.

In addition to being not tight for general degree constraints, the polymatroid
bound has another disadvantage: the linear program~\eqref{eqn:polymatroid:bound}
has an exponential size in query complexity. While this is ``acceptable'' in
theory, it is simply not acceptable in practice. Typical OLAP queries we have
seen at {\sf LogicBlox} or {\sf RelationalAI} have on average $20$ variables; and
$2^{20}$ certainly cannot be considered a ``constant'' factor, let alone
analytic and machine learning workloads which have hundreds if not thousands of
variables. We next present a sufficient condition allowing for the polymatroid
bound to not only be tight, but also computable in polynomial time in query
complexity.

\bdefn[Acyclic degree constraints]
Associate a directed graph $G_\dc = ([n], E)$ to the degree constraints $\dc$ by
adding to $E$ all directed edges $(x,y) \in X \times (Y-X)$ for every
$(X,Y,N_{Y|X}) \in \dc$. If $G_\dc$ is acyclic, then $\dc$ is said to be {\em
acyclic degree constraints}, in which case any topological ordering (or linear
ordering) of $[n]$ is said to be {\em compatible} with $\dc$.
The graph $G_\dc$ is called the {\em constraint dependency graph} associated
with $\dc$.
\edefn

Note that if there are only cardinality constraints, then $G_\dc$ is empty and
thus $\dc$ is acyclic. In particular, acyclicity of the constraints does not 
imply acyclicity of the query, and the cardinality constraints do not affect
the acyclicity of the degree constraints.
In a typical OLAP query, if in addition to cardinality constraints we have FD
constraints including non-circular key-foreign key lookups, then $\dc$ is
acyclic.
Also,
verifying if $\dc$ is acyclic can be done efficiently in $\poly(n,|\dc|)$-time.

\bprop
Let $Q$ be a query with acyclic degree constraints $\dc$;
then the following hold:
\begin{equation}
   \max_{h \in \Mod_n \cap \hdc} h([n])=
   \max_{h \in \overline\Gamma^*_n \cap \hdc} h([n])=
   \max_{h \in \Gamma_n \cap \hdc} h([n]).
   \label{eqn:all:equal}
\end{equation}
In particular, the polymatroid bound is tight and computable in $\poly(n,|\dc|)$-time.
\label{prop:collapse}
\eprop
\bp
Let $h^*$ denote an optimal solution to the linear program
$\max \{ h([n]) \suchthat h \in \Gamma_n \cap \hdc\}$.
Because $\Mod_n \subseteq \overline\Gamma^*_n \subseteq \Gamma_n$,
to prove~\eqref{eqn:all:equal} it is sufficient to exhibit a {\em modular} function
$f \in \Mod_n \cap \hdc$ for which $f([n]) = h^*([n])$.

Without loss of generality, assume the identity permutation is compatible with
$\dc$, i.e. for every $(X,Y,N_{Y|X}) \in \dc$, we have $x < y$ for all $x \in X$
and $y \in Y-X$. Define a set function $f : 2^{[n]} \to \R_+$ as follows:
\begin{equation}
   f(S) := \begin{cases}
      0 & \text{ if }S = \emptyset\\
      h^*([i])-h^*([i-1]) & \text{ if }S = \{i\}, i \in [n]\\
      \sum_{i \in S} f(i)
      & \text{ if }S \subseteq [n], |S|>1.
   \end{cases}
\end{equation}
The function $f$ is clearly modular because $h^*$ is monotone.
The fact that $f([n]) = h^*([n])$ follows from the telescoping sum.
It remains to show that $f \in \hdc$. 
We will show by induction on $|Y-X|$
that $f(Y|X) \leq h^*(Y|X)$ for any degree constraint
$(X,Y,N_{Y|X}) \in \dc$.
The base case when $Y=X$ holds trivially.
Let $(X,Y, N_{Y|X})$ be any degree constraint in $\dc$ where $|Y-X|>0$. 
Let $j$ be the largest integer in $Y-X$. We have
\begin{align}
   f(Y \suchthat X) 
%   &= f(j)+f(Y-\{j\})-f(X)\\
   &= h^*([j] \suchthat [j-1])+f(Y-\{j\} \suchthat X)\\
   &\leq h^*([j] \suchthat [j-1])+h^*(Y-\{j\} \suchthat X)
   \label{eqn:induction:hypothesis}\\
   &= h^*([j-1] \cup Y \suchthat [j-1])+h^*(Y-\{j\} \suchthat X)\\
   &\leq h^*(Y \suchthat Y \cap [j-1])+h^*(Y-\{j\} \suchthat X)
   \label{eqn:sm:h*}\\
   &= h^*(Y \suchthat Y - \{j\})+h^*(Y-\{j\} \suchthat X)\\
   &= h^*(Y \suchthat X)\\
   &\leq \log N_{Y|X}. \label{eqn:last:one}
\end{align}
Inequality~\eqref{eqn:induction:hypothesis} follows from the induction
hypothesis,~\eqref{eqn:sm:h*} from submodularity of $h^*$,
and~\eqref{eqn:last:one} from the fact that $h^* \in \hdc$.

Lastly, the polymatroid bound is computable in $\poly(n,|\dc|)$-time
because the linear program $\max \{ h([n]) \suchthat h \in \Mod_n \cap \hdc\}$
has polynomial size in $n$ and $|\dc|$. To see this, define a variable
$v_i=h(i)$ for every $i \in [n]$, then the modular LP is 
\begin{align}
   \max        &\quad \sum_{i=1}^n v_i \label{eqn:vertex:packing}\\
   \text{s.t.} &\quad \sum_{i \in Y-X} v_i \leq \log_2 N_{Y|X} &&
   (X,Y,N_{Y|X})\in \dc\\
   &\quad v_i \geq 0 && \forall i \in [n].
\end{align}
\ep

Associate a dual variable $\delta_{Y|X}$ for every $(X,Y,N_{Y|X}) \in \dc$.
In what follows for brevity we sometimes write $(X,Y) \in \dc$ instead of
the lengthier $(X,Y, N_{Y|X}) \in \dc$.
The dual LP of~\eqref{eqn:vertex:packing} is the following
\begin{align}
   \min        &\quad \sum_{(X,Y,N_{Y|X}) \in \dc} \delta_{Y|X}\log_2 N_{Y|X}
   \label{eqn:edge:covering}\\
   \text{s.t.} &\quad \sum_{\substack{(X,Y) \in \dc \\ i \in Y-X}} \delta_{Y|X}
   \geq 1 && \forall i \in [n] \label{eqn:covering:i}\\
   &\quad \delta_{Y|X} \geq 0 && \forall (X,Y) \in \dc.
\end{align}
This is exactly $\agm$-bound if $\dc$ contains only cardinality constraints,
and hence our proposition is a generalization of $\agm$-bound and its tightness.

\section{Algorithms}
\label{sec:algos}

An algorithm evaluating $Q(\bm D)$ under degree constraints $\dc$ is a $\wcoj$
algorithm if it runs in time 
\[ \tilde O\left( |\bm D| +
2^{\max_{h \in \overline\Gamma^*_n \cap \hdc} h([n])}
\right).
\]
In general, we do not know how to even compute the entropic bound, in part
because there is no finite set of linear inequalities characterizing the
entropic cone~\cite{matus2007infinitely}. 
Hence, thus far we have settled for designing algorithms meeting the polymatroid
bound, running in time
$\tilde O\left( |\bm D| +
2^{\max_{h \in \Gamma_n \cap \hdc} h([n])}
\right)$. This question is difficult enough, and in some cases (e.g.
Proposition~\ref{prop:collapse}), the two bounds collapse.
We we present two such algorithms in this section, the first algorithm is
inspired by the proof of Friedgut's inequality, and the second is guided by a
proof of a particular type of information theoretic inequalities called
Shannon-flow inequalities.

\subsection{An algorithm for acyclic degree constraints}

For simplicity, we first assume that there is a variable order compatible with
$\dc$, and w.l.o.g. we assume the order is $(1,\dots, n)$.
Algorithm~\ref{algo:backtracking} is a backtracking search algorithm inspired by
the inductive proof of Theorem~\ref{thm:friedgut}. Our analysis is summarized in
the following theorem. Note that the runtime expression
does not hide any factor behind $\tilde O$.

\floatname{algorithm}{Procedure}
\renewcommand{\algorithmicrequire}{\textbf{Input: }}
\renewcommand{\algorithmicensure}{\textbf{Output: }}

\begin{algorithm}[t]
   \algorithmicrequire{Query $Q$, acyclic degree constraints $\dc$ \\
   \qquad \qquad $(1,\dots,n)$ compatible with $\dc$
   }\;
   \SetKwFunction{proc}{search}
   \Return \proc{$()$}; \tcp*[f]{empty-tuple argument}\;
   \SetKwProg{myproc}{SubRoutine}{}{}
   \myproc{\proc{$\bm a_S$}}{
      $i \leftarrow |S|+1$\;
      \If {$i > n$} {
         \Return $\bm a_S$\;
      } \Else {
         $P \leftarrow \emptyset$\;
         \For {$\displaystyle{a_i \in \bigcap_{\substack{(X,Y) \in \dc \text{ s.t.  }i \in
               Y-X\\ R \text{ guards } (X,Y)}} \pi_{A_i} \sigma_{\bm A_{S\cap
               Y}=\bm a_{S\cap Y}} \pi_Y R}$} {
                  $P \leftarrow P \bigcup \proc{$(\bm a_S, a_i)$}$\;
         }
         \Return $P$\;
      }
   }
   \caption{Backtracking Search for Acyclic $\dc$}
   \label{algo:backtracking}
\end{algorithm}

\bthm
Let $Q$ be a query with acyclic degree constraints $\dc$. Suppose $(1,\dots,n)$ is
compatible with $\dc$. 
Let $\bm D\models \dc$ be a database instance. 
Then,
Algorithm~\ref{algo:backtracking} runs in worst-case optimal time:
\begin{equation}
   O\left(
   n\cdot |\dc| \cdot \log |\bm D| \left[
      |\bm D| + 2^{\max_{h\in \overline\Gamma^*_n\cap \hdc} h([n])}
      \right]
   \right)
\end{equation}
\ethm
\bp
Let $\bm \delta$ denote an {\em optimal} solution to the
LP~\eqref{eqn:edge:covering}. Then, by Proposition~\ref{prop:collapse} and strong
duality of linear programming, it is sufficient to show that
Algorithm~\ref{algo:backtracking} runs in time
\begin{equation}
   O\left(
   n\cdot |\dc| \cdot \log |\bm D| \left[
      |\bm D| + \prod_{(X,Y,N_{Y|X}) \in \dc} N_{Y|X}^{\delta_{Y|X}}
      \right]
   \right).
   \label{eqn:sufficient:runtime}
\end{equation}
The $O(n \cdot |\dc| \cdot |\bm D|\log |\bm D|)$ term in~\eqref{eqn:sufficient:runtime} comes
from a preprocessing step where we precompute and index the projections
$\pi_YR$ in the algorithm.
We show the remaining runtime by induction on $n$. When $n=1$, the only thing the 
algorithm does is compute the intersection
\begin{equation}
   I := \bigcap_{\substack{(\emptyset,Y, N_{Y|\emptyset}) \in \dc \text{ s.t.  }1 \in
      Y\\ R \text{ guards } (\emptyset,Y, N_{Y|\emptyset})}} \pi_{A_1} R.
      \label{eqn:I}
\end{equation}
The intersection can be computed in time proportional to the smallest set, up to
a $\log |\bm D|$ factor. And thus, up to a $\log |\bm D|$ factor,
the runtime is
\begin{align*}
   \min_{\substack{(\emptyset,Y,N_{Y|\emptyset}) \in \dc \text{ s.t.  }1 \in Y\\ R \text{ guards
   } (\emptyset,Y,N_{Y|\emptyset})}} |\pi_{A_1} R|
   &\leq 
   \prod_{\substack{(\emptyset,Y,N_{Y|\emptyset}) \in \dc \text{ s.t.  }1 \in Y\\ R \text{ guards }
   (\emptyset,Y,N_{Y|\emptyset})}} |\pi_{A_1}  R|^{\delta_{Y|\emptyset}}\\
   &\leq 
   \prod_{(X,Y,N_{Y|X}) \in \dc} N_{Y|X}^{\delta_{Y|X}}.
\end{align*}
The first inequality follows because the minimum of a set of non-negative reals
is bounded above by their geometric mean, and from the fact that $\bm \delta$
satisfies~\eqref{eqn:covering:i}.

When $n>1$, the algorithm implicitly or explicitly computes $I$
in~\eqref{eqn:I}, which can be done within the budget time as shown in the base
case. Then, for each binding $a_1 \in I$, Algorithm~\ref{algo:backtracking}
performs backtracking search on the remaining variables $(A_2,\dots,A_n)$. By
induction, up to a $\log |\bm D|$ factor, this can be done in  time
\begin{align*}
   &
   \sum_{a_1 \in I} 
   \prod_{\substack{(\emptyset,Y, N_{Y|\emptyset}) \in \dc\\ 1 \in Y\\
   R \text{ guards } (\emptyset,Y,N_{Y|\emptyset})}}
   |\sigma_{A_1=a_1}
   \pi_YR|^{\delta_{Y|\emptyset}}
   \cdot 
   \prod_{\substack{(X,Y, N_{Y|X}) \in \dc\\ 1 \notin Y-X}}
   N_{Y|X}^{\delta_{Y|X}}\\
   &\leq
   \prod_{\substack{(\emptyset,Y, N_{Y|\emptyset}) \in \dc\\ 1 \in Y\\ 
   R \text{ guards } (\emptyset,Y,N_{Y|\emptyset})}}
   \left(
   \sum_{a_1 \in I} 
   |\sigma_{A_1=a_1}
   \pi_YR|
   \right)^{\delta_{Y|\emptyset}}
   \cdot 
   \prod_{\substack{(X,Y, N_{Y|X}) \in \dc\\ 1 \notin Y-X}}
   N_{Y|X}^{\delta_{Y|X}}\\
   &\leq
   \prod_{(X,Y,N_{Y|X})\in\dc} N_{Y|X}^{\delta_{Y|X}}.
\end{align*}
The first inequality follows from Theorem~\ref{thm:friedgut} and the fact that
$\bm\delta$ satisfies~\eqref{eqn:covering:i}.
The second inequality follows from the fact that $R$ guards
$(\emptyset,Y,N_{Y|\emptyset})$.
\ep

Algorithm~\ref{algo:backtracking} has several key advantages:
(1) It is worst-case optimal when $\dc$ is acyclic;
(2) It is very simple and does not require any extra memory (after
pre-processing): we can iterate through the output tuples without
computing intermediate results;
(3) It is friendly to both hash or sort-merge strategies, as the only required
assumption is that we can compute set intersection in time proportional to the
smallest set.

What if $\dc$ is {\em not} acyclic? There are two solutions. The first solution
is to find an
acyclic collection $\dc'$ of degree constraints giving the 
smallest worst-case output size bound and run Algorithm~\ref{algo:backtracking} on $\dc'$.
(The final output is semijoin-reduced against the guards of the original degree
constraints $\dc$.)
The second solution is to run the more general and more sophisticated algorithm called
$\panda$ we will present in Section~\ref{subsec:panda}.

We discuss in this section more details regarding the first strategy of
constructing an acyclic $\dc'$.
How do we know that such an acyclic $\dc'$ exists and is satisfied by the input
database? And, how do we know that the corresponding worst-case output size
bound is {\em finite}?
For example, the first thought that comes to mind may be to try to remove one or
more constraints from $\dc$ to make it acyclic. However, this na\"ive strategy
may result in an infinite output size bound. Consider the following query
\begin{equation}
   Q(A,B,C,D) \leftarrow R(A), S(A,B), T(B,C), W(C,A,D).
\end{equation}
The degree constraints given to us are $N_{A|\emptyset}$ guarded by $R$, 
$N_{B|A}$ guarded by $S$,
$N_{C|B}$ guarded by $T$, and
$N_{AD|C}$ guarded by $W$. 
In particular we do not know the sizes of $S$, $T$, and $W$. (They can be
user-defined functions/relations, which need not be materialized.)
It is easy to see that removing any of the input constraints will yield an
infinite output size bound. 
(An infinite output size bound corresponds precisely
to situations where some output variable is unbound and cannot be inferred from
bound variables by chasing $\FD$s.)

\bprop
Let $Q$ be a full conjunctive query with
degree constraints $\dc$ such that $\sup_{\bm D \models \dc}Q(\bm D)$ is finite.
Then, there exists an {\em acyclic} set of degree constraints $\dc'$ for which
\bi
 \item[(i)] For any database instance $\bm D$, 
    if $\bm D \models \dc$ then $\bm D \models \dc'$.
 \item[(ii)] $\sup_{\bm D\models \dc'}Q(\bm D)$ is finite.
\ei
\label{prop:acyclic:modification}
\eprop
\bp
Let $V$ be the set of all variables occuring in $Q$. 
We define the set of {\em bound} variables recursively as follows.
For any constraint $(X, Y, N_{Y|X})$, if all variables in $X$ are bound, then
all variables in $Y$ are also bound.
In particular, the cardinality constraint $(\emptyset,
Y, N) \in \dc$ implies that all variables in $Y$ are bound.
We make the following two claims.

{\bf Claim 1.} $\sup_{\bm D \models \dc} |Q(\bm D)|$ is finite if and only if 
all variables in $V$ are bound.

{\bf Claim 2.} If $\dc$ is cyclic, then for any cycle $C$ in $G_\dc$ 
there is a variable $y \in C$ on the cycle for which the following holds.
There is a
constraint $(X,Y,N_{Y|X})$ with $y \in Y-X$ such that, if we replace
$(X,Y,N_{Y|X})$ by $(X, Y-\{y\}, N_{Y-\{y\}|X} := N_{Y_X})$, then all variables
in $V$ remain bound under the new set of constraints.

The two claims prove the proposition, because if $\dc$ is still cyclic, we can
apply the above constraint replacement to obtain a new degree constraint set
$\dc'$ with $\sup_{\bm D \models \dc'}Q(\bm D)$ remains finite. Any relation $R$
guarding the degree constraint $(X,Y,N_{Y|X})$ is still a guard for the new
degree constraint $(X,Y-\{y\},N_{Y-\{y\}|X})$. 
Hence, $\bm D \models \dc$ implies $\bm D \models \dc'$.
We can repeat this process until $\dc'$ reaches acyclicity.
We next prove the two claims.

{\em Proof of Claim 1.}
For the forward directly, suppose $\sup_{\bm D\models \dc}|Q(\bm D)|$ is finite.
Let $B$ denote the set of bound variables, and $U$ denote the set of unbound
variables. 
Assume to the contrary that $U \neq \emptyset$. 
Then, for any degree constraint $(X,Y,N_{Y|X})\in \dc$ we have $X \cap U =
\emptyset$ implies $Y \cap U = \emptyset$ also.
Let $h \in \overline\Gamma^*_n \cap \hdc$ denote an arbitrary entropic function. 
Let $c>0$ be an arbitrary constant.
Define a new set function $f : 2^{[n]} \to \R_+$ by 
\begin{align}
   f(S) := \begin{cases} h(S) & S \cap U = \emptyset\\
      h(S)+c & S \cap U \neq \emptyset.
   \end{cases}
\end{align}
Then, we can verify that $f \in \overline\Gamma^*_n \cap \hdc$ as well. 
First, $f \in \overline\Gamma^*_n$ because
it is a non-negative linear combination of two entropic
functions\footnote{Recall that $\overline\Gamma^*_n$ is a convex cone.}.
Second, $f \in \hdc$ because $f(Y|X) = h(Y|X)$ for every constraints 
$(X,Y,N_{Y|X})\in \dc$: 
   \begin{align}
      f(Y|X) &= f(Y)-f(X) = \begin{cases}
         h(Y)-h(X) = h(Y|X) & X \cap U = \emptyset\\
         h(Y)+c-(h(X)+c) = h(Y|X) & X \cap U \neq \emptyset.
      \end{cases}
   \end{align}
Note that $f(V) = h(V)+c$; and,
since $c$ was arbitrary, $\sup_{\bm D \models \dc} |Q(\bm D)| = \max_{h \in
\overline\Gamma^*_n\cap \hdc} h(V)$ is unbounded.

Conversely, assume every variable is bound. Since $h(V) \leq \sum_v h(v)$ for
every $h \in \overline\Gamma^*_n$, it is
sufficient to show that $h(v)$ is finite for every $h \in \overline\Gamma^*_n
\cap \hdc$. There must exist some cardinality constraint $(\emptyset, Y,
N_{Y|\emptyset})$ in order for all variables to be bound. Then, $h(y) \leq
h(Y) \leq 
\log N_{Y|\emptyset}$ for all $y \in Y$. Inductively, consider a degree constraint
$(X, Y, N_{Y|X})$ for which $h(x)$ is finite for all $x \in X$, then for any $y
\in Y$ we have $h(y) \leq h(Y) \leq h(X)+\log N_{Y|X})$.

{\em Proof of Claim 2.}
A sequence 
$(X_1=\emptyset, Y_1, N_1),
(X_2, Y_2, N_2),
\cdots,
(X_k, Y_k, N_k)$ 
of constraints is said to {\em reach} a vertex $v \in V$
if the following holds: for any $i \in [k]$, $X_i \subseteq
\bigcup_{j=1}^{i-1}Y_j$, and $v \in  Y_k$.
From Claim 1, there is a sequence of degree constraints reaching every 
variable in $V$.
Consider the shortest sequence of $k$ degree constraints reaching some vertex 
$y \in C$. Then, there is no vertex of $C$ in the set
$Y_1 \cup \cdots \cup Y_{k-1}$.
And thus, because $X_i \subseteq \bigcup_{j=1}^{i-1} Y_j$,
there is no vertex of $C$ in the set $X_1 \cup \cdots \cup X_k$ either.
Now, let $(X, Y, N_{Y|X})$ denote a degree constraint for which $(x,y)$ is on
the cycle $C$, and $(x,y) \in X \times (Y-X)$. Then $(X,Y,N_{Y|X})$ is not part
of the degree constraint sequence. Consequently, when we turn $(X, Y, N_{Y|X})$
into $(X, Y-\{y\}, N_{Y-\{y\}|X}=N_{Y|X})$ all vertices of $V$ are still
bound because the constraint change can only affect the boundedness of $y$, and
$y$ can still be reached via the degree constraint sequence.
\ep

The proof of the above proposition also suggests a simple brute-force algorithm
for finding the best acyclic constraint set $\dc'$. The algorithm runs in
exponential time in query complexity. 
It also raises a natural question: when is the worst-case output size on
the best acyclic $\dc'$ {\em the same} as that of $\dc$?
We do not know the general answer to this question; however, there is one case
when the answer is easy. 
Recall that a {\em simple $\FD$} is an $\FD$ of the form $A_i \to A_j$ for two
single variables $A_i$ and $A_j$.
The following implies a result from Gottlob et al.~\cite{GLVV}.

\bcor
If $\dc$ contains only cardinality constraints and simple $\FD$s, then in
polynomial time in query complexity, we can determine a subset $\dc' \subseteq
\dc$ so that $\dc'$ is acyclic and, more importantly, 
$\sup_{\bm D\models \dc}Q(\bm D) = \sup_{\bm D\models \dc'}Q(\bm D)$.
In particular, Algorithm~\ref{algo:backtracking} is a $\wcoj$ algorithm for $Q$.
\ecor
\bp
The constraints in the set $\hdc$ are either cardinality constraints of the form
$h(Y) \leq N_{Y|\emptyset}$ or $\FD$-constraints of the form $h(\{i,j\}) =
h(\{i\})$. Since equalities are transitive, if there was a cycle in $G_\dc$ we
can remove one edge from the cycle without changing the feasible region defining
$\hdc$. Keep breaking directed cycles this way, we end up with the acyclic
$\dc'$ as desired.
\ep

\subsection{$\panda$}
\label{subsec:panda}

Finally we informally present the main ideas behind the $\panda$
algorithm~\cite{panda},
which can achieve the polymatroid-bound runtime, modulo huge polylog and
query-dependent factors. The algorithm actually solves a particular form of 
{\em disjunctive datalog rules}, of which conjunctive queries are a special
case. Most importantly, it leads to algorithms meeting highly refined notions of
``width parameters'' over tree decompositions of the query.
In summary, $\panda$ has far reaching theoretical implications
in terms of the class of problems it helps solve and the insights it provides in
designing and reasoning about join algorithms; at the same time, the hidden
query-dependent and polylog in the data factors leave much room for desire.
Materials in this section are exclusively from~\cite{panda}, with some
simplification.

\subsubsection{Understanding the polymatroid bound}

The starting point of designing any algorithm meeting the polymatroid bound is
to understand in detail what the bound entails.
To this end, we establish some notations.
For any $I, J \subseteq [n]$, we write $I \incomp J$ to mean $I \not\subseteq J$
{\em and} $J \not\subseteq I$.
In order to avoid rewriting $\log_2 N_{Y|X}$ and $(X,Y,N_{Y|X})\in \dc$
over and over, we define 
\begin{align}
   \calP &:= \{(X,Y) \suchthat \emptyset \subseteq X \subsetneq Y \subseteq [n]\}\\
   n_{Y|X} &:= \begin{cases}
      \log_2 N_{Y|X} & (X,Y,N_{Y|X}) \in \dc\\
      +\infty & \text{otherwise}.
   \end{cases}
\end{align}
Note that $|\calP| = \sum_{i=0}^n \binom n i (2^i-1)=3^n-2^n$, and
the vector $\bm n := (n_{Y|X})$ lies in $\R_+^{\calP}$.
The polymatroid bound $\max_{h \in \Gamma_n \cap \hdc}h([n])$
is the optimal objective value of
the following optimization problem:
\begin{align}
   \max & & h([n]) \label{eqn:first:primal:lp}\\
   \text{s.t.} & &h(Y)- h(X)  &\leq n_{Y|X}, & (X,Y) \in \calP \nonumber\\
                    & &h(I\cup J)+h(I\cap J)-h(I) - h(J) &\leq 0,       & I \incomp J \nonumber\\
                    & &h(Y)-h(X)                   &\geq 0,       & (X,Y) \in
                    \calP\nonumber
\end{align}
In addition to the degree constraints $h(Y)-h(X) \leq n_{Y|X}$, signifying $h
\in \hdc$, the remaining constraints spell out the definition of a polymatroid:
submodularity, monotonicity, and non-negativity, where monotonicity and 
non-negativity collapsed into one constraint $h(Y|X) \geq 0$.
It is thus more notationally convenient to work on the space 
$\bm h = (h(Y|X)) \in \R_+^\calP$ instead of the original space of functions $h
: 2^{[n]} \to \R_+$. 

\bdefn[Conditional polymatroids]
We refer to the vectors $\bm h = (h(Y|X)) \in \R_+^\calP$ that are feasible to the last two
constraints of~\eqref{eqn:first:primal:lp} and to~\eqref{eqn:dual:comp:decomp}
below as the {\em conditional polymatroids}.
Also, we will write $h(Y)$ instead of $h(Y|\emptyset)$ for brevity.
\edefn
The equivalent linear program in the space of conditional polymatroids is
\begin{align}
   \max & & h([n]|\emptyset) \label{eqn:primal:lp}\\
   \text{s.t.} & &h(Y|X)  &\leq n_{Y|X}, & (X,Y) \in \calP \nonumber\\
                    & &h(I\cup J | J) - h(I | I\cap J) &\leq 0,       & I
                    \incomp J \label{eqn:dual:subm}\\
                    & &h(Y|X)+h(X|\emptyset)-h(Y|\emptyset) &= 0, & (X,Y) \in
                    \calP\label{eqn:dual:comp:decomp}\\
                    & &h(Y | X)                   &\geq 0,       & (X,Y) \in \calP\nonumber
\end{align}
In the above, when performing the space transformation from polymatroids to
conditional polymatroids, we impose the obvious extra ``conservation''
constraints $h(Y|\emptyset) = h(Y|X)+h(X|\emptyset)$. 
Note that this implies (and thus is equivalent to) the more general conservation
constraints $h(Y|Z) = h(Y|X)+h(X|Z)$. 

The (primal) linear program is very clean, but it does not give us a clear sense
of the bound relative to the input statistics $n_{Y|X}$. To obtain this
relationship, we look at the dual linear program.
Associate dual variables $\bm \delta = (\delta_{Y|X})$ to the degree constraints,
$\bm\xi = (\xi_{I,J})$ to the submodularity constraints, 
$\bm\alpha = (\alpha_{X,Y})$ to the extra conservation constraints, 
then the dual LP can be written as follows.
\begin{align}
   \min        && \inner{\bm \delta, \bm n} \label{eqn:dual:LP}\\
   \text{s.t.} &&\flow(\emptyset, [n]) & \geq 1, & \nonumber\\
               &&\flow(X,Y) & \geq 0, & (X,Y) \in \calP \wedge (X,Y) \neq (\emptyset, [n]),\nonumber\\
               &&(\bm \delta, \bm \xi) &\geq\bm 0.\nonumber
\end{align}
where for any $(X,Y) \in \calP$, $\flow(X,Y)$ is defined by
\begin{align*}
   \flow(\emptyset, Y) &:= \delta_{Y|\emptyset} - \sum_{X : (X,Y) \in \calP}
   \alpha_{X,Y} + \sum_{W : (Y,W) \in \calP} \alpha_{Y,W}
   - \sum_{\substack{J : Y\incomp J\\Y\cap J = \emptyset}}\xi_{Y,J}\\
   \flow(X,Y) &:= 
   \delta_{Y|X} +
   \sum_{\substack{I : I\incomp X\\I\cup X = Y}}\xi_{I,X}
   - \sum_{\substack{J : Y\incomp J\\Y\cap J = X}}\xi_{Y,J}
      + \alpha_{X,Y} && X \neq \emptyset
\end{align*}
Note that there is no non-negativity requirement on $\bm\alpha$.
The dual LP~\eqref{eqn:dual:LP} is important in two ways. First, let $\bm h^*$
and $(\bm \delta^*, \bm\xi^*,\bm\alpha^*)$ denote a pair of primal- and dual-optimal
solutions, then from strong duality of linear programming~\cite{MR88m:90090} we
have 
\begin{equation}
   h^*([n]) = \inner{\bm\delta^*,\bm n} = \sum_{(X,Y)\in\calP}
\delta^*_{Y|X}n_{Y|X}.
   \label{eqn:strong:duality}
\end{equation}
We are now able to ``see'' the input statistics
contributions to the objective function.
(The reader is welcome to compare this expression with that
of~\eqref{eqn:edge:covering}.)
Second, the dual formulation allows us to formulate a (vast) generalization of
Shearer's inequality, to deal with general degree constraints. We refer to this
generalization as ``Shannon-flow inequalities'', which is discussed next.

\subsubsection{Shannon-flow inequalities}

\bdefn[Shannon-flow inequality]
Let $\bm\delta \in \R_+^\calP$ denote a non-negative 
coefficient vector. If the inequality
\begin{equation}
   h([n]) \leq \inner{\bm \delta, \bm h}
   \label{eqn:sfi}
\end{equation}
holds for all conditional polymatroids $\bm h$, then it is called a {\em Shannon-flow
inequality}.
\edefn
It is important not to lose sight of the fact that conditional polymatroids are
simply a syntactical shortcut to the underlying
polymatroids, designed to simplify notations; they are not a new function class.
We can rewrite~\eqref{eqn:sfi} in a wordier form involving only polymatroids:
$h([n]) \leq \sum_{(X,Y) \in \calP} \delta_{Y|X} (h(Y) - h(X)).$

Shannon-flow inequalities occur naturally in characterizing feasible solutions to
the dual LP~\eqref{eqn:dual:LP} using Farkas' lemma~\cite{MR88m:90090}:

\bprop[From~\cite{panda}]
Let $\bm\delta \in \R_+^\calP$ denote a non-negative 
coefficient vector. The inequality $h([n]) \leq \inner{\bm\delta,\bm h}$ is a
Shannon-flow inequality {\em if and only if} there exist $\bm\xi,
\bm\alpha$
such that $(\bm\delta,\bm\xi,\bm\alpha)$ is a feasible solution to the dual
LP~\eqref{eqn:dual:LP}.
\label{prop:sfi}
\eprop

It is not hard to show that Shearer's inequality is a consequence.

\bcor[Shearer's inequality]
Let $\calH = ([n], \calE)$ be a hypergraph, and $\bm\delta = (\delta_F)_{F\in
\calE}$ be a vector of non-negative coefficients. Then, the inequality
$h([n]) \leq \sum_{F \in \calE} \delta_F h(F)$ holds for all polymatroids iff
$\bm \delta$ is a fractional edge cover of $\calH$.
\ecor

\subsubsection{Proof sequences and $\panda$}

We next explain how studying Shannon-flow inequalities leads to an algorithm
called the {\bf P}roof {\bf A}ssisted e{\bf N}tropic {\bf D}egree {\bf Aware}
($\panda$) algorithm whose runtime meets the polymatroid bound.
At a high level, the algorithm consists of the following steps.
\be
   \item Obtain an optimal solution $(\bm\delta^*,\bm\xi^*,\bm\alpha^*)$ to the
      dual LP~\eqref{eqn:dual:LP}. 
   \item Use the dual solution to derive a particular mathematical proof of the
      Shannon-flow inequality $h([n]) \leq \inner{\bm\delta^*,\bm h}$. This
      inequality is a Shannon-flow inequality thanks to
      Proposition~\ref{prop:sfi}. This mathematical proof has to be of a
      particular form, called the ``proof sequence,'' which we define below.
   \item Finally, every step in the proof sequence is interpreted as a symbolic
      instruction to perform a relational operator (partition some relation, or
      join two relations). These symbolic instructions are sufficient to compute
      the final result.
\ee
The full version of $\panda$ is more complex than the three steps above, due to
several technical hurdles we have to overcome. The reader is referred
to~\cite{panda} for the details. This section can only present a 
simplified high-level structure of the algorithm.

We next explain the proof sequence notion in some details.
If we were to expand out a
Shannon-flow inequality $h([n]) \leq \inner{\bm\delta^*,\bm h}$, it would be of
the form
$h([n]) \leq \sum_{(X,Y)\in \calP} \delta^*_{Y|X} h(Y|X),$
where $\delta^*_{Y|X}$ are all non-negative. Hence, we interpret the RHS of the
above inequality as a set of ``weighted'' conditional polymatroid terms:
the term $h(Y|X)$ is weighted by $\delta^*_{Y|X}$.
To prove the inequality, one may attempt to apply either the submodularity
inequality~\eqref{eqn:dual:subm} or the equality~\eqref{eqn:dual:comp:decomp} to
some of the terms to start converting the RHS to the LHS.
These applications lead to three types of rules:
\bi
 \item Suppose $\delta \cdot h(I | I \cap J)$ occurs on the RHS, then we may
apply~\eqref{eqn:dual:subm} with a weight of $w \in [0,\delta]$ to obtain a (new or not)
weighted term $w \cdot h(I\cup J | J)$ while retain $(\delta-w) \cdot h(I | I \cap J)$ of the
old term. In this scenario, we say that we have applied the 
\emph{submodularity rule} 
$
   h(I | I \cap J) \to h(I\cup J | J)
$
with a weight of $w$.
 \item The equality~\eqref{eqn:dual:comp:decomp} can be used in two ways: either
    we replace $h(Y|\emptyset)$ by $h(Y|X)+h(X|\emptyset)$, resulting in a 
    \emph{decomposition rule} 
    $
    h(Y|\emptyset) \to h(Y|X)+h(X|\emptyset), 
    $
    or the other way around where we'd get a 
    \emph{composition rule} 
    $
    h(Y|X)+h(X|\emptyset) \to h(Y).
    $
    These rules can be applied with a weight, as before.
\ei
A (weighted) {\em proof sequence} of a Shannon-flow inequality is a series of weighted
rules such that at no point in time any weight is negative, and that in the end
$h([n])$ occurs with a weight of at least $1$.
%The {\em length} of the proof sequence is the number of rules in it.
We were able to prove the following result:

\bthm[From~\cite{panda}]
There exists a proof sequence for every Shannon-flow inequality $h([n]) \leq
\inner{\bm\delta,\bm h}$%, whose length is a function of $n$ and $\bm\delta$.
\ethm

The dual feasible solution $(\bm \delta, \bm\xi,\bm\alpha)$ gives us an
intuition already on which rule to apply. For example, $\xi_{I,J}>0$ indicates
that we should apply the submodularity rule, $\alpha_{X,Y}>0$ indicates a
decomposition rule, and $\alpha_{X,Y}<0$ hints at a composition rule. The
technical issue we have to solve is to make sure that these steps {\em
serialize} to a legitimate proof sequence.

%Note also that the dual feasibility of 
%$(\bm \delta, \bm\xi,\bm\alpha)$ has nothing to do with the input statistics,
%only the objective function is! In particular, the length of our proof sequence
%is independent of the input statistics.

Finally, after obtaining the proof sequence for the inequality $h([n]) \leq
\inner{\bm \delta^*,\bm h}$, $\panda$ interprets the proof sequence as follows.
Note that $\delta^*_{Y|X}>0$ implies $n_{Y|X}<\infty$, which means there is a
guard for the corresponding constraint. We associate the guarding relation with
the term $h(Y|X)$.
Now, we look at each rule in turn:
a decomposition rule corresponds to partitioning the relation associated with
the conditional polymatroid term being decomposed;
a composition rule corresponds to joining the two associated relations;
and a submodularity rule is used to move the association map.
These concepts are best illustrated with an example, which is a minor
modification of an example from~\cite{csma}.

{\small
\begin{table*}[th!]
   \begin{tabular}{l|l|l|l}
      \hline
      Name & proof step & operation & action\\
      \hline
      \hline
   {decomposition} & $h(BC) \to h(B)+h(BC|B)$ & partition& $S \to S^\heavy \cup S^\light$ \\
      &&& $S^{\heavy} \leftarrow \{(b,c) \in S \ : \ |\sigma_{B=b}R| > \theta\}$\\
      &&& $S^{\light} \leftarrow \{(b,c) \in S \ : \ |\sigma_{B=b}R| \leq \theta\}$\\
      \hline
      {submodularity} &$h(CD) \to h(BCD|B)$ & {\sf NOOP} & $T(C,D)$ now ``affiliated'' with $h(BCD|B)$ \\
      \hline
      {composition} &$h(B)+h(BCD|B) \to h(BCD)$& join & $I_1(B,C,D) \leftarrow S^{\heavy}(B,C), T(C,D)$.  \\
      \hline
   {submodularity} & $h(ABD|BD) \to h(ABCD|BCD)$ & {\sf NOOP} & $V(A,B,D)$ now ``affiliated'' with $h(ABCD|BCD)$ \\
      \hline
   {composition} &$h(ABCD|BCD)+h(BCD) \to h(ABCD)$ & join & $\text{\sf output}_1(A,B,C,D) \leftarrow V(A,B,D), I_1(B,C,D)$.  \\
      \hline
      {submodularity} &$h(BC|B) \to h(ABC|AB)$& {\sf NOOP} & $S^\light$ now ``affiliated'' with $h(ABC|AB)$ \\
      \hline
      {composition} &$h(AB)+h(ABC|AB) \to h(ABC)$& join & $I_2(A,B,C) \leftarrow R(A,B), S^\light(B,C)$.  \\
      \hline
      {submodularity} &$h(ACD|AC) \to h(ABCD|ABC)$& {\sf NOOP} & $W(A,C,D)$ now ``affiliated'' with $h(ABCD|ABC)$ \\
      \hline
      {composition} &$h(ABC)+h(ABCD|ABC) \to h(ABCD)$ & join & $\text{\sf
      output}_2(A,B,C,D) \leftarrow I_2(A,B,C), W(A,C,D)$.\\
      \hline
      \hline
   \end{tabular}
   \caption{Proof sequence to algorithmic steps. $\theta :=
   \sqrt{\frac{N_{BC}N_{CD}N_{ABD|BD}}{N_{AB}N_{ACD|AC}}}$}
   \label{table:algo}
\end{table*}
}

\begin{ex}
Consider the following query
\begin{equation*}
   Q(A,B,C,D) \leftarrow R(A,B), S(B,C), T(C,D),
   W(A,C,D), V(A,B,D),
\end{equation*}
with the following degree constraints:
\bi
\item $(\emptyset, AB, N_{AB})$ guarded by $R$,
\item $(\emptyset, BC, N_{BC})$ guarded by $S$,
\item $(\emptyset, CD, N_{CD})$ guarded by $T$,
\item $(AC, ACD, N_{ACD|AC})$ guarded by $W$,
\item $(BD, ABD, N_{ABD|BD})$ guarded by $V$.
\ei
We claim that the following is a Shannon-flow inequality:
\begin{equation*}
   h(ABCD) \leq \frac 1 2 [
      h(AB)+h(BC)+h(CD)+h(ACD|AC)+h(ABD|BD)
   ],
   \label{eqn:sample:sfi}
\end{equation*}
and $\panda$ can evaluate the query in time
\begin{equation} 
   \tilde O\left( \sqrt{N_{BC}N_{CD}N_{ABD|BD} N_{AB} N_{ACD|AC}} \right). 
   \label{eqn:alloted:time}
\end{equation}
%   For example, if $N_{AB}=N_{BC}=N_{CD}=N$, and
%   $N_{ABD|BD}=N_{ACD|AC}=1$, i.e. $BD \to A$ and $AC\to D$
%   were functional dependencies, then $\panda$ evaluates the query in time
%   $\tilde O(N^{3/2})$. 

Inequality~\eqref{eqn:sample:sfi} holds for every polymatroid $h \in \Gamma_4$, because 
\begin{align*}
   &   h(AB)+h(BC)+h(CD)+h(ACD|AC)+h(ABD|BD)\\
   &=  h(AB)+h(B)+h(BC|B)+h(CD)+h(ACD|AC)+h(ABD|BD)\\
   &\geq  h(AB)+h(B)+h(BC|B)+h(BCD|B)+h(ACD|AC)+h(ABD|BD)\\
   &=  h(AB)+h(BC|B)+h(BCD)+h(ACD|AC)+h(ABD|BD)\\
   &\geq  h(AB)+h(BC|B)+h(BCD)+h(ACD|AC)+h(ABCD|BCD)\\
   &=  h(AB)+h(BC|B)+h(ACD|AC)+h(ABCD)\\
   &\geq  h(AB)+h(ABC|AB)+h(ACD|AC)+h(ABCD)\\
   &= h(ABC)+h(ACD|AC)+h(ABCD)\\
   &\geq h(ABC)+h(ABCD|ABC)+h(ABCD)\\
   &= h(ABCD)+h(ABCD).
\end{align*}
The proof above applied the proof sequence shown in Table~\ref{table:algo},
which also contains the step-by-step description of how to 
translate the proof sequence into an algorithm.
The total runtime is within~\eqref{eqn:alloted:time}:
\begin{equation}
   \tilde O\left(\frac{N_{BC}}{\theta}N_{CD}N_{ABD|BD} +
   \theta N_{AB} N_{ACD|AC}\right)
   = \tilde O\left(
   \sqrt{N_{BC}N_{CD}N_{ABD|BD} N_{AB} N_{ACD|AC}}
   \right).
\end{equation}
\end{ex}

\section{Open problems}
\label{sec:opm}

There are many interesting and challenging
open questions arising from this line of inquiries:
questions regarding the bounds, the algorithms, the desire to make them
practical and extend their reach to more difficult or realistic settings.
In terms of bounds, the most obvious question is the following:
\begin{opm}
   Is the entropic bound computable?
\end{opm}
We know that the entropic bound is tight, but we do not know if it is decidable
whether the bound is below a given threshold. This question is closely
related to the question of determining whether a linear inequality is satisfied
by all entropic functions or not. 

Next, assuming we have to settle for the polymatroid bound, then the 
challenge is to find efficient algorithms for computing the polymatroid bound,
which is a linear program with an exponential number of variables.
As we have seen, there are classes of queries and degree constraints for which
we can compute the polymatroid bound in polynomial time in query copmlexity.
Even if computing the polymatroid bound is difficult in general, say it is $\np$-hard or
harder, it would be nice to be able to characterize larger classes of queries and
constraints allowing for tractability.

\begin{opm}
   What is the computational (query) complexity of computing the polymatroid
   bound? Design an efficient algorithm computing it.
\end{opm}

Another important line of research is to enlarge the class of degree constraints
for which the polymatroid bound is tight, making $\panda$ a $\wcoj$
algorithm (up to a large polylog factor).
In these special cases, perhaps there are simpler algorithms than $\panda$,
such as Algorithm~\ref{algo:backtracking}.
The following two questions are along this direction.

\begin{opm}
   Characterize the class of queries and degree constraints for which the best
   constraint modification as dictated by
   Proposition~\ref{prop:acyclic:modification} has the same worst-case output
   size bound as the original constraint set.
\end{opm}

\begin{opm}
   Characterize the class of degree constraints $\dc$ for which the polymatroid
   bound is tight.
\end{opm}

$\panda$ is a neat algorithm, which is capaable of answering the more general problem of
evaluating a disjunctive datalog rule. Hence, perhaps there are faster
algorithms without the large polylog factor,
designed specifically for answering conjunctive queries:

\begin{opm}
   Find an algorithm running within the polymatroid bound that does not impose
   the poly-log (data) factor as in $\panda$.
\end{opm}

Reasoning about entropic inequalities has allowed us to gain deeper insights on
both the algorithm design and bounding the worst-case output size.
Entropy is, by definition, an expectation. And thus it should serve as a
bridge to reasoning about {\em average} output size.
A result from Atserias et al.~\cite{AGM} which has not been exploited further by
the database community is their concentration result, a good starting point for
the following question.

\begin{opm}
   Develop a theory and algorithms for average-case output size bound.
\end{opm}

Average case bounds and complexity is only one way to go {\em beyond worst-case}.
Another line of research is on the notion of {\em instance-optimality} for
computing joins.
Instance optimality is a difficult notion to define formally, let alone having
an optimal algorithm under such stringent requirement. 
There are only a few known work on instance-optimality in database 
theory~\cite{minesweeper,grades15,tetris,DBLP:conf/pods/FaginLN01}.
After information theory perhaps geometric ideas will play a
bigger role in answering this question:

\begin{opm}
   Develop a theory and practical algorithms for instance-optimal query
   evaluation.
\end{opm}

Last but not least, traditional database optimizers have been designed on the
``one join at a time'' paradigm, influenced by relational algebra
operators. It is this author's strong belief that the time is ripe for the
theory and practice of multiway join optimizers, based on information
theoretic analysis and sampling strategies, taking into account systems 
requirements such as streaming, transactional constraints, incremental 
view maintenance, etc.

\begin{opm}
   Develop a theory and practical algorithms for an optimizer for the multiway
   join operator.
\end{opm}

%Database, both theory and practice, should not shy away from addressing more
%difficult join problems. The assumption that ``most practical DB queries are
%acyclic'' is a restraining assumption. It restrains database theory and
%algorithms to address a more interesting class of problems, it restrains
%database system builders from building systems capable of supporting
%constraint satisfication and machine learning applications, expanding the
%boundaries and potentials of an RDBMS.

%\bibliographystyle{siam}
%\bibliography{main}

\end{document}